\documentclass[aps,prb,reprint,groupedaddress, showkeys]{revtex4-1}

\usepackage{times}
\usepackage[utf8]{inputenc}
\usepackage{amsmath}
\usepackage{amssymb}
\usepackage{graphicx}
\usepackage{bm}
\DeclareBoldMathCommand{\bfM}{{\sf M}}
\usepackage{subfigure}
\usepackage{multirow}
\usepackage{dsfont}
\usepackage[pdftex,table]{xcolor}

\begin{document}
\sloppy

\title{Remarks about surface plasmons and their stability}

\author{Gino Wegner and Carsten Henkel}
\email[]{henkel@uni-potsdam.de}
\affiliation{Institute of Physics and Astronomy,
University of Potsdam, Germany}

\date{\today}

\begin{abstract}
We comment on the macroscopic model for surface plasmons
of H.-Y. Deng [\emph{New J.\ Phys.}\ {\bf 21} (2019) 043055; arXiv:1712.06101]
and a claim, based on energy conversion from charges to the electric field,
that surface plasmons on metallic
surfaces may become unstable
[\emph{J. Phys.:\ Cond.\ Matt.}\ {\bf 29} (2017) 455002; arXiv:1606.06239, 1701.01060].
The discussion revolves around the formulation of charge conservation in the bulk
and the surface of a metal. We elaborate in particular on the role of
a finite electric current normal to the surface.
Using a scheme of Cercignani \& Lampis and of Zaremba,
we point out that the model chosen by Deng
for the non-specular scattering of electrons
needs to be amended to prevent the disappearance of charges at the surface.
Different models and approaches in the literature on
surface plasmons are reviewed: 
the interfacial excess field approach of Bedeaux and Vlieger
which contains Deng's macroscopic model,
the assumption of specular reflection of Ritchie and Marusak,
a hydrodynamic model with a composite charge density (partially localized
at the surface), 
the local dielectric model,
and a macroscopic method with (anti)symmetric fictitious stimuli 
(used, e.g., by Garc\'{i}a-Moliner and Flores).
This puts Deng's results into perspective and illustrates problems with
his approach.
\end{abstract}


\maketitle

In this Comment, we would like to address a series of papers by
H.-Y. Deng on surface plasmons at metal-dielectric interfaces.
The first one appeared in 2015 on the arXiv (with co-workers
K. {W}akabayashi and C.-H. {L}am)
and was finally published in \emph{Phys.\ Rev.\ B} 
under the title
``Universal self-amplification channel for surface plasma waves''
[Ref.\,\onlinecite{Universal_Self_Amplification_Deng_et_al_2017}]. 
A paper claiming the same phenomenon to occur in a metallic film appeared 
in the same year \cite{Deng_2017b_metal_film},
followed by 
an alternative argument for the instability of surface plasma waves
based on energy conservation
\cite{Possible_Instability_Deng_2017}.
In \emph{New J. Phys.} (2019), Deng has also addressed 
the question why this prediction went unnoticed in the 
literature.\cite{Deng_2019}
In other arXiv posts \cite{2017arXiv170603404D,2018arXiv180608308D,Deng_2020a},
the instability of surface plasmons is also mentioned, while 
the scope is widened, e.g., to energy electron loss spectroscopy.

The discovery of the surface plasmon dates back to 1957 when Ritchie \cite{Ritchie_Plasma_Losses} formulated a general model for the energy loss 
spectrum of charged particles passing through a thin metal foil, a topic that
had attracted numerous experimental investigations.
Ritchie could successfully describe collective and individual excitations 
of the metallic electrons, that are accompanied by oscillating electric fields.
 Next to the already known concept of volume plasmons introduced by Pines and Bohm \cite{PhysRev.82.625,PhysRev.85.338}, Ritchie found 
an additional loss peak below the plasma frequency which is proper to a
bounded metal and was later called the surface plasmon. 
In the 1970s the theoretical description of (surface) plasmons
moved from classical grounds to incorporate quantum aspects of the electron
response \cite{Plummer_et_al_impact_surface_plasmon,Raimes_1957},
using, e.g., the jellium model to describe the metal and
applying tools of the ideal Fermi gas \cite{Mukhopadhyay_1978, Apell_1978}.
Theoretical and experimental progress has led to a variety of applications
in the now sprawling field of plasmonics, for example nano-scale light
sources as the surface plasmon nanolaser \cite{app9050861} 
or spaser \cite{Bergman_2003,Spaser_biological_probe} 
that may be useful as a biosensor, in microscopy, optical computing 
and information storage.
A big challenge to real-life plasmonic devices are the large losses in metals. 
Dissipation channels are provided by collisions of conduction electrons, 
by Landau damping and interband absorption. 
This can be mitigated by the introduction of amplifying media 
(see, e.g., Refs.\,\onlinecite{Bergman_2003,Smuk_2006}).
In contrast to these proposals, Deng has suggested that there exist 
an ``intrinsic channel of amplification'' that would involve the
ballistic motion of carriers reflected from the metal surface\cite{Universal_Self_Amplification_Deng_et_al_2017, Possible_Instability_Deng_2017, 2018arXiv180608308D, Deng_2019}.
This has triggered the present work. We consider a geometry similar to
Deng's, consisting of a metal-vacuum interface that is planar on the 
macroscopic scale, the metallic body being essentially infinitely thick.
The conduction electrons are described by the jellium model (i.e., ignoring
the structure of the crystal backbone).

To make the paper self-contained, we recall in Sec.\,\ref{sec:Deng_general_model_surface}
the main ideas of Deng's model. We analyze his statement that
a non-vanishing normal current must exist at the surface to warrant the existence
of surface plasmons. A key issue is how to deal with charge conservation and
the spatial structure of the charge distribution. 
In Sec.\,\ref{sec:SCM}, we present in detail the semiclassical model 
that is used for the
electronic response, based on the Boltzmann equation
in the relaxation time approximation and supplemented by
boundary conditions. We review the history of 
scattering models (partially specular, partially diffuse) and compare to Deng's results.
{A detailed estimate of the surface plasmon loss rate, based on
the energy balance argument of Ref.\,\onlinecite{Possible_Instability_Deng_2017}, 
is computed in 
Sec.\,\ref{sec:Calculation_amplification_loss_rate}. 
We find indeed, similar to Deng, that the scattering of electrons at the metal
surface produces one term corresponding to amplification.
The surface plasmon is, however, overall lossy when all terms are taken into account. 
A surface contribution to the energy balance crucial to Deng's analysis 
is argued to be questionable.}
In Sec.\,\ref{sec:From_HDM_to_LDM} we show how a macroscopic electrodynamic
model can be 
embedded into an approach based on excess interfacial fields developed by
Bedeaux and Vlieger (BV). 
This clarifies the issue whether charge conservation provides sufficient information
to calculate the dispersion relation. 
We also recall that the usual hydrodynamic model with a vanishing normal surface current 
does indeed allow for surface plasmons. 
At this stage, a mathematical and physical discussion on how to perform the local limit 
is given.
In Sec.\,\ref{sec:pseudo_model}, we supplement the macroscopic descriptions 
of Deng and of BV by a surface electrodynamics approach
due to Garc\'{i}a-Moliner and Flores (GF)\cite{GMF_cond_Surfaces_non_specular, GMF_intro_theory_solid_surfaces},
designed with an apparently similar scope to Deng's 
recent work\cite{Deng_2019}. For the response in the bulk metal, one could take here
any conductivity with or without spatial dispersion. 
Following GF, we display
the surface plasmon dispersion for a hydrodynamic model.
Sec.\,\ref{sec:conclusion} summarizes
our conclusions about the validity of Deng's proposal.

\section{\label{sec:Deng_general_model_surface}A charge-centered formulation of surface electrodynamics}

\subsection{Geometry: metallic half-space}
\label{ss:half-space}

The geometry considered is that of a metal and a dielectric (typically vacuum)
occupying the half-spaces $z \ge 0$ and $z < 0$, respectively. 
The metal surface is macroscopically
located at $z = 0$ and infinitely extended in the $xy$-plane. It appears flat
on the scale of the plasmon wavelength $2\pi / k$ where $k$ is the wave vector
parallel to the surface. The description of 
Refs.\,\onlinecite{Universal_Self_Amplification_Deng_et_al_2017, Possible_Instability_Deng_2017, Deng_2019} is macroscopic in the sense
that the microscopic details in the surface region (its width {$d_{\rm s}$} is
typically a few lattice constants) are not resolved: we work in the limit
{$k d_{\rm s} \to 0$}. The electric current density in the metal can then be written as
\begin{equation}
\bm{j}(\bm{x}, t) = \bm{J}(\bm{x}, t) \Theta(z)
 \,,
\label{eq:complete_surface_description_half space_geometry}
\end{equation}
where $\bm{J}$ is the current inside the metal at position $\bm{x} = (x,y,z)^{\sf T}$
and $\Theta$ the Heaviside step function that represents the rapid change 
in the surface region. 

Deng's formulation of the surface plasmon problem is focused on the dynamics
of the charge and current densities, while the electric field is eliminated
in a self-consistent way by using Coulomb's law. His formulation of charge 
conservation\cite{Possible_Instability_Deng_2017,Deng_2019} 
takes the form
\begin{equation}
\mbox{Deng:}\quad
\left( \partial_t + \tau^{-1}\right) \rho
+ \nabla \cdot \bm{j} = 0\,,
\label{eq:wrong_formulation_continuity}
\end{equation}
where $1/\tau$ is the collision rate of conduction electrons. 
Because of the collision term, Eq.\,(\ref{eq:wrong_formulation_continuity})
does not locally conserve charge. Actually, it
follows from a naive application
of the relaxation time approximation to the Boltzmann equation 
(see Sec.\,\ref{sec:SCM} below). 
The failure of not conserving charge locally should be discussed carefully.
In the history of metal optics, this problem appeared several times in different 
disguise -- a brief summary is given in 
Subsec.\,\ref{subsec:charge-conservation}.
We may ignore it for the moment, knowing that most of Deng's actual
estimates are taken in the collisionless limit $\tau \to \infty$ anyway.

Inserting Eq.\,(\ref{eq:complete_surface_description_half space_geometry}) for the
current density, one gets for $z \ge 0$
\begin{equation}
\mbox{Deng:}\quad
\left( \partial_t + \tau^{-1}\right) \rho
+ \Theta(z)\nabla \cdot \bm{J} 
=
- \Theta'(z) J_z( \bm{x}_0 )
\,,
\label{eq:continuity-with-delta-source}
\end{equation}
where $\bm{x}_0 = (x, y, 0)$ is a position in the surface
and
$\Theta'(z) = \delta(z)$ is localized in the surface region.
The term $-\Theta'(z) J_z( \bm{x}_0 )$ on the right hand side 
is described by Deng
[\onlinecite[after Eq.\,(24)]{Universal_Self_Amplification_Deng_et_al_2017}]:
``Physically, the right hand side of Eq.\,(\ref{eq:continuity-with-delta-source}) 
means that charges must pile up on the surface if they do not come to a halt 
before they reach it.'' 
He considers this term
to be crucial for the existence of a surface plasmon which
is claimed to exist only if the normal component $J_z( \bm{x}_0 ) = 
\lim_{z \downarrow 0} J_z(\bm{x})$ of the 
current density (called ``surface current'' in the following) does not 
vanish.
Otherwise, ``the surface would be completely severed from the rest of the metal''~\cite{Possible_Instability_Deng_2017}, 
and only volume plasmons could be excited.\cite{Deng_2019}

According to an anecdote told by Plummer 
et al.\cite{Plummer_et_al_impact_surface_plasmon}, 
a discussion between Ritchie and Gabor brought up the insight 
that it is actually the electric field normal to the surface that must 
be nonzero to generate a surface plasmon. This condition does not 
imply that the surface current be nonzero: examples are provided
by various formulations of a non-local current response (spatial dispersion)
that go back historically to the anomalous skin effect 
(see Secs.\,\ref{sec:SCM} and\,\ref{sec:From_HDM_to_LDM}).

On the microscopic scale, the current smoothly changes between the bulk values in 
both media, vacuum and metal in our case. But on a macroscopic scale, information 
about the surface region gets lost. One key question
that we would like to clarify here is what are the length scales 
a given ``macroscopic'' model actually tries to resolve (or not). 
Different cases disagree among each other in terms of the values (zero or not) 
of surface current and charge.
In this spirit, 
the localized term $\Theta'(z)$ on the right of
Eq.\,(\ref{eq:continuity-with-delta-source}) 
is the result of not resolving the surface region in which the
electron density drops to zero.

To illustrate this point, consider the Drude model for the response of the
metallic electrons to the electric field. 
We shall treat the system in linear response and assume that all fields are
proportional to $\exp {\rm i} (k x - \omega t)$. 
Only a $z$-dependence remains. We also work on 
spatial scales where electromagnetic retardation can be neglected.
The electric field,
for example, can then be generated by a potential $\phi( z )$ and has nonzero 
components $E_x = -{\rm i} k \phi$ and $E_z = - \partial_z \phi$. 

In the Drude model, the current density $\bm{J}( z ) = \sigma(\omega) \bm{E}( z )$
depends only on the local electric field at the same position in the metal
($z \ge 0$).
By taking the integral of Eq.\,(\ref{eq:wrong_formulation_continuity})
or~(\ref{eq:continuity-with-delta-source}) 
over a thin layer centered around $z = 0$ whose thickness eventually shrinks 
to zero, we get
\begin{equation}
- {\rm i} \omega \rho_{\rm s} + J_z( 0^+ ) = 0
\label{eq:continuity-charge-surface-current}
\end{equation}
This links the time derivative of the charge $\rho_{\rm s}$ in the surface
layer and the normal component of the current density, evaluated in the metal
just outside the layer. (On the vacuum side, $j_z(0^-) = 0$, of course.)
Note that we have adopted here the canonical 
formulation of the continuity equation, dropping the relaxation time $\tau$
from Eq.(\ref{eq:wrong_formulation_continuity}).
By Coulomb's law, the surface charge generates an electric potential
(cgs units)
\begin{equation}
\phi( z ) = 2\pi \rho_{\rm s} \, \frac{ {\rm e}^{- k |z| } }{ k }
\label{eq:potential-surface-charge}
\end{equation}
whose decay length $1/k$ is set by the periodic variation of all fields 
parallel to the surface. According to the Drude model, the surface current
in Eq.\,(\ref{eq:continuity-charge-surface-current}) takes the form
\begin{equation}
J_z( 0^+ ) = k \sigma( \omega ) \phi( 0 ) 
= k \frac{ {\rm i} \omega_{\rm p}^2 }{ 4\pi ( \omega + {\rm i} / \tau) } \phi( 0 ) 
\label{eq:surface-current-Drude}
\end{equation}
where $\omega_{\rm p}$ is the plasma frequency and $\tau$ the relaxation time
of the electric current (average scattering time of electrons). 
The three 
equations~(\ref{eq:continuity-charge-surface-current})--(\ref{eq:surface-current-Drude})
yield, provided that $\rho_{\rm s} \ne 0$, the dispersion equation
$\omega \bar{\omega} = \omega_{\rm p}^2 / 2$ with $\bar{\omega} = \omega + {\rm i} / \tau$.
Its solution is
\begin{equation}
\omega(k) = \sqrt{ 
	\frac{ \omega_{\rm p}^2 }{ 2 }
	-
	\frac{ 1 }{ 4 \tau^2 } 
	}
- \frac{ {\rm i} }{ 2\tau }
\qquad
\mbox{(Drude model)}
\label{eq:Drude-dispersion-relation}
\end{equation}
This simple calculation provides a starting point to compare 
with Deng's results regarding several points. 

The imaginary part of the surface plasmon frequency
satisfies $-1/\tau < \mathop{\rm Im} \omega(k) < 0$:
the surface plasmon is damped. For the auxiliary complex frequency
$\bar{\omega}$, one gets $\mathop{\rm Im} \bar{\omega} \ge 0$, as claimed
by Deng. He works with a relaxation term in the continuity equation 
[see Eq.\,(\ref{eq:wrong_formulation_continuity})] and finds a local
dispersion equation in the form $\bar{\omega}^2 = \omega_{\rm p}^2 / 2$,
giving a damping twice as large. 
While in Eq.\,(\ref{eq:Drude-dispersion-relation}), $\mathop{\rm Im} \bar{\omega}$
depends on the relaxation rate $1/\tau$, the text
in Ref.\,\onlinecite{Universal_Self_Amplification_Deng_et_al_2017},
around Eq.\,(8),
claims the contrary.

For typical metals, we have
$\omega_{\rm p} \gg 1/\tau$ so that 
the frequency $\omega_{\rm s} = \omega_{\rm p} / \sqrt{2}$ sets the 
long-wavelength limit of the surface plasmon dispersion (real part)
at a metal--vacuum interface.
Several authors have shown, both on general grounds and for particular models,
that this remains true beyond the 
local (Drude) approximation.\cite{Wagner_Oberfl_Wellen_e_Plasma, Ritchie_Marusak,Feibelman_1971,Flores_1972,Zaremba_PhysRevB.9.1277}
Deng claims in Refs.\,\onlinecite{Universal_Self_Amplification_Deng_et_al_2017, Possible_Instability_Deng_2017} that the long-wavelength limit 
$\omega( k \to 0 )$ should depend on model parameters 
[see Eq.\,(\ref{eq:Deng-dispersion-0}) below].

The calculation assumes that the charge density is nonzero only in the thin surface layer
$z \sim 0$, otherwise the electric potential~(\ref{eq:potential-surface-charge})
must be modified. In the local approximation,
the electric field component $E_z$ changes sign when $z = 0$ is crossed.
This may have led to an erroneous result that can be inferred from 
Ref.\,\onlinecite{Possible_Instability_Deng_2017}
[the paragraph after Eq.\,(11)]:
there $J_z(0^+)$
appears with a sign opposite to Eq.\,(\ref{eq:surface-current-Drude}). 

The passage to the local limit and the handling of charge conservation in general,
will be a key point of our discussion in what follows. For example,
we analyze in Sec.\,\ref{sec:usual-HDM-BC} how the results of the hydrodynamic model
recover those of the Drude approach. In Appendix~\ref{as:cos-and-exp-integrands},
we give a discussion of integral representations for the electric field, 
in particular how they behave in the limit $z \downarrow 0$.

\subsection{\label{sec:Deng_universal_macro_description}Plasmon dispersion relation as an eigenvalue problem}

{%
We come back to Deng's formulation of the surface plasmon where a non-local
relation between current and field (spatially dispersive conductivity) is adopted. 
This motivates the introduction of an 
integral operator $\hat{\mathcal{H}}$ that relates the current divergence
and the charge according to
}
\begin{equation}
\left(- {\rm i}\bar{\omega} \right) \nabla\cdot\bm{J}(z) 
= \hat{\mathcal{H}}\rho(z) 
= \int dz'~\mathcal{H}(z,z')\rho(z')
\,,
\label{eq:Deng_surface_model_continuity_with_integral_operator}
\end{equation}
where again $\bar{\omega} = \omega + {\rm i} / \tau$.
As an intermediate step, 
the operator $\hat{\mathcal{H}}$ involves solving the Poisson equation
to get the electric potential for a given charge density.
For the conductivity, a semiclassical
kinetic theory based on the Boltzmann equation in the relaxation time approximation similar to earlier work by 
Reuter and Sondheimer\cite{Reuter_Sondheimer_anomalous_Skineffect} 
and Wagner\cite{Wagner_Oberfl_Wellen_e_Plasma}, for example, is taken. 
This calculation is simplified
by representing the charge density as a cosine transform:
\begin{equation}
\rho(z) = \frac{2}{\pi} \int_0^{\infty}\!dq~\rho_q \cos(qz) \,.
\label{eq:definition_cosine_transform}
\end{equation}
In addition, Deng focuses on that part of the operator $\hat{\mathcal{H}}$ 
that describes 
the excitations of an infinite system, neglecting terms depending
on scattering at the surface. (For a discussion of this approximation, 
refer to the paragraph after Eq.\,(\ref{eq:Deng_decomposition_conductivity}).)
Its (double) cosine transform may then be given by
\begin{equation}
\hat{\mathcal{H}}(q,q') \approx \Omega^2(k, q) \delta(q-q')
\, ,
\label{eq:Deng_macroscopic_response_volume_operator_eigenvalue_Fourier}
\end{equation}
where $\Omega( k, q)$ is related to the bulk dispersion relation. 
Using this result 
in the cosine transform of Eq.\,(\ref{eq:continuity-with-delta-source})
and using
Eq.\,(\ref{eq:Deng_surface_model_continuity_with_integral_operator}),
one finds that the charge density is given by
\begin{equation}
\rho_q  = \frac{{\rm i}\bar{\omega} J_z(0)}{\Omega^2(k, q) - \bar{\omega}^2}
\,,
\label{eq:Deng_macro_description_charge_by_surf_curr}
\end{equation}
where $J_z(0)$ is the amplitude
of the surface current. (If the correct continuity equation is used, 
the denominator contains the product $\omega \bar{\omega}$ rather 
than $\bar{\omega}^2$.)
If one sets $J_z(0) = 0$ and seeks a solution with 
$\rho_q \ne 0$, then Eq.\,(\ref{eq:Deng_macro_description_charge_by_surf_curr}) 
yields the dispersion relation for bulk excitations, 
$\bar{\omega}^2 - \Omega^2(k, q) = 0$ (neglecting, as mentioned, the
influence of boundary conditions on the bulk plasmon spectral density 
\cite{Barton_1979}).

The last step towards surface plasmon modes is to express the surface 
current by another integral operator 
\begin{equation}
{\rm i}\bar{\omega} J_z(0) = \int_0^{\infty}\! dq~\frac{G(\bm{K}, \omega)}{k^2+q^2} \rho_q 
\,,
\label{eq:def-G-kernel}
\end{equation}
where $\bm{K} = (k, 0, q)$.
If we insert Eq.\,(\ref{eq:Deng_macro_description_charge_by_surf_curr}) for
$\rho_q$ under the integral and simplify both sides by $J_z(0) \ne 0$,
we get
\begin{equation}
\int_0^{\infty}\! \frac{dq}{k^2+q^2} 
\frac{G(\bm{K}, \omega)}{\Omega^2(k,q) - \bar{\omega}^2} = 1
\,.
\label{eq:Deng_def_surface_diel_fctn}
\end{equation}
Deng solves this equation numerically. The results can be written
as a complex dispersion relation for surface 
plasmons \cite{Universal_Self_Amplification_Deng_et_al_2017, Possible_Instability_Deng_2017}
\begin{equation}
\omega = \omega_s(k) + {\rm i} (\gamma_0(k) - 1/\tau)
\,,
\label{eq:Deng-dispersion-0}
\end{equation}
where Fig.\,1 from 
Ref.\,\onlinecite{Universal_Self_Amplification_Deng_et_al_2017}
gives the following typical values:
for $k \approx 0.07 \,\omega_{p} / v_{\rm F}$, the surface plasmon
frequency is
$\omega_s(k) \sim 0.9\,\omega_{\rm p}$
and its imaginary part $\gamma_0(k) \sim 0.1 \ldots 0.2 \,\omega_{\rm p}$,
a positive value highlighting the instability. 
(Recall that for metals like gold and 
silver, one has $\omega_{\rm p} \sim 100/\tau$.)
Here $\omega_{\rm p}$ is the metal's plasma 
frequency and $v_{\rm F}$ the Fermi velocity.
In Ref.\,\onlinecite{Possible_Instability_Deng_2017},
the approximation 
\(
\gamma_0(k) \approx \omega_{\rm p} (  0.16 - 0.066 \,p)
- 0.25\, k v_{\rm F}
\)
is found,
where $0 \le p \le 1$ is a parameter describing the fraction of
electrons that show specular reflection at the metal surface.

It is remarkable that these numbers deviate strongly from earlier work.
The long-wavelength limit of the real part $\omega_s(k) \to \omega_{\rm p} / \sqrt{2}$
(for a metal-vacuum interface) is well 
established, follows from the matching
of macroscopic fields in the local approximation (dielectric function
$\varepsilon( \omega ) = - 1$) and is consistently recovered in models including 
spatial 
dispersion\cite{Wagner_Oberfl_Wellen_e_Plasma,Ritchie_Marusak,Feibelman_1971,Zaremba_PhysRevB.9.1277}.
The imaginary part of the dispersion relation
is even more surprising: it is the key claim of Deng's 
papers\cite{Universal_Self_Amplification_Deng_et_al_2017,Possible_Instability_Deng_2017,Deng_2019}
that the ballistic motion of electrons, after reflection from the surface,
provides an \emph{amplification} channel that may overtake the loss rate
$1/\tau$ in Eq.\,(\ref{eq:Deng-dispersion-0}).
One may raise the question why
in that case the Fermi sea of filled electronic levels should become 
unstable\cite{Possible_Instability_Deng_2017},
since it is constructed as the state of lowest energy for a fixed charge density.

\subsection{\label{sec:Deng_macro_model_approximations}Approximations within Deng's  description}

From Eqs.\,(\ref{eq:Deng_surface_model_continuity_with_integral_operator},
\ref{eq:def-G-kernel})
we learn that the kernels $\mathcal{H}$ and ${G}$ can be determined 
from the current density which is itself given by
\begin{equation}
J_{\mu}(z) = \sum\limits_{\nu} 
\int dz'\, \tilde{\sigma}_{\mu\nu}(z,z') E_{\nu}(z')
\label{eq:lin_resp_current}
\end{equation}
{within linear response theory. Here, $\mu, \nu \in\{x,y,z\}$ label Cartesian
components. Exploiting the translation symmetry of our surface problem, the
conductivity tensor $\tilde{\sigma}_{\mu\nu}(z,z')$ depends on two positions
in addition to wavevector
$k$ and frequency $\omega$ (spatial and temporal dispersion). The integration
over the variable $z'$ translates for example the ballistic
motion of electrons on the scale of the mean free path $v_{\rm F} \tau$.
The dependence of $\tilde{\sigma}$ on two positions $z$, $z'$ (rather than
their difference) describes
the breaking of translational symmetry by the boundary conditions.}
We emphasize that $\tilde{\sigma}(z,z')$ is the conductivity tensor
of the metallic half-space and includes surface scattering.
It can be generally decomposed into
\begin{equation}
\tilde{\sigma}(z,z') = \sigma_{\rm b}(z-z') + \sigma_{\rm s}(z,z')\,,
\label{eq:Deng_decomposition_conductivity}
\end{equation}
where $\sigma_{\rm b}$ is the bulk conductivity
and $\sigma_{\rm s}$ embodies surface effects. Deng passes this decomposition 
on to the operators $\mathcal{H}$ and $G$.

There are two assumptions that lead to the diagonalization of $\mathcal{H}$
in Eq.\,(\ref{eq:Deng_macroscopic_response_volume_operator_eigenvalue_Fourier}): 
First, Deng assumes that the current in the metallic bulk is mainly determined
by $\sigma_{\rm b}$ and neglects the contribution of $\sigma_{\rm s}$ which
is expected to be localized at the surface. We discuss in Sec.\,\ref{sec:pseudo_model}
an approach which takes $\sigma_{\rm s}$ into account.
Second, the cosine transform that is used to represent the charge density
in Eq.\,(\ref{eq:definition_cosine_transform}), 
makes boundary terms vanish that involve the derivative $\partial_z \rho(0)$.
This is not true, however, for a charge distribution that behaves like
$\rho(0) \exp( - \kappa z )$, for example, whose cosine transform is simply
\begin{equation}
\rho_q = \frac{ \rho(0)\, \kappa }{ q^2 + \kappa^2 }
\,.
\label{eq:cosine-transform-exp-charge}
\end{equation}
We show in Secs.\ref{sec:BV_formalism_application_to_HDM},
\ref{sec:Deng-critique-HDM} for a hydrodynamic
model that a finite gradient $\partial_z \rho(0)$ appears naturally because of
spatial dispersion; it can be
conveniently represented within a Fourier expansion of $\rho(z)$. In this model,
a surface plasmon is found although the surface current vanishes, $J_z(0) = 0$.
It provides a counter-example to Deng's interpretation of 
Eq.\,(\ref{eq:Deng_macro_description_charge_by_surf_curr}) that surface plasmons
with a finite charge density should necessarily have a nonzero surface current.

The decomposition of the kernel $G$ in Eq.\,(\ref{eq:def-G-kernel}) into
bulk and surface parts $G_{\rm b} + G_{\rm s}$ follows from
Eqs.\,(\ref{eq:lin_resp_current}, \ref{eq:Deng_decomposition_conductivity}),
evaluating the current at $z = 0$. Deng identifies 
the surface part $G_{\rm s}$ with translation symmetry breaking and finds it
to play a key role for the amplification of the surface plasmon. 
In all explicit applications {except when using the Boltzmann equation 
(semiclassical model, SCM)}, he approximates the bulk conductivity $\sigma_{\rm b}$ 
by its local Drude form for the calculation of $G_{\rm b}$.

In contrast to Deng's proposition, the specular reflection model does, in fact, 
include translation symmetry breaking. This can be seen by writing the
integral\,(\ref{eq:lin_resp_current}) in the form
\begin{eqnarray}
&& J_\mu(z) = \int_0^\infty\!dz' 
\sum_\nu\sigma_{{\rm b},\mu\nu}( z - z' ) E_\nu( z' )
\label{eq:translationally_variant_SRM_conductivity}
\\
&& + \int_0^\infty\!dz'\Big[
\sigma_{{\rm b},\mu x}( z + z' ) E_x( z' )
-
\sigma_{{\rm b},\mu z}( z + z' ) E_z( z' )
\Big]
\,.
\nonumber
\end{eqnarray}
The first term alone would be the result of the so-called dielectric 
approximation\cite{Heinrichs_PhysRevB.7.3487}
where the integration range in Eq.\,(\ref{eq:lin_resp_current}) is restricted to
$z' \ge 0$ and $\tilde\sigma$ is replaced by its bulk version. This approximation
is also used by Deng\cite{Deng_2019} when he computes the approximate form
$\mathcal{H}_{\rm b}$ to get Eq.\,(11). (This can be seen from the lower integration
limit of the $z''$-integral written after his Eq.\,(6).)
The second term describes the specular scattering of charges accelerated towards 
the surface. Following Ritchie and Marusak\cite{Ritchie_Marusak}, 
it can be viewed as the response
of the bulk conductivity to the electric field extended by mirror symmetry to $z' < 0$
\begin{equation}
\bm{E}(-z') = \bfM \bm{E}(z'>0)~,
\label{eq:SRM_mirror_symmetric_electric_field}
\end{equation}
where $\bfM = \mathop{\rm diag}( 1, 1, -1 )$ is the mirror reflection
at the surface $z = 0$.
{Such an extended field has been called ``pseudo-field'' in the work
of Garc\'{i}a-Moliner and Flores, and the sign flip of its normal component
corresponds to a ``fictitious'' charge sheet at $z = 0$.
From Eq.\,(\ref{eq:translationally_variant_SRM_conductivity}), 
the decomposition\,(\ref{eq:Deng_decomposition_conductivity}) of the
conductivity can be read off, and obviously leads, in the limit $z \to 0$, 
to a finite value for 
$G_{\rm s}$, i.e., translation symmetry breaking is present.}

The approximations outlined here illustrate that there are implicit
additional boundary conditions behind the approach of Deng which seems difficult
to be considered universal. For the boundary conditions used in Deng's
version of the semiclassical model, see Sec.\,\ref{sec:SCM}.

\subsection{Charge conservation}
\label{subsec:charge-conservation}

{Another} reason for the deviation of Deng's results from the literature
may be found in the way charge conservation is handled.
It was already pointed out that 
Eq.\,(\ref{eq:wrong_formulation_continuity}) does not conserve charge in the
metal bulk 
because of the scattering rate $1/\tau$. This problem arises from the relaxation
time approximation to the Boltzmann equation that guarantees charge conservation
only at the global level. As suggested by Mermin\cite{Mermin_PhysRevB.1.2362}, 
this can be improved by specifying that the collision term relaxes the system's
distribution function to its local equilibrium value. 
In Sec.\,\ref{sec:SCM}, we show how this
correction can be used to recover the actual form 
of the continuity equation.
Another way to enforce local charge conservation is to use the
Boltzmann equation only
to determine the current density and to solve for $\rho$ from the continuity 
equation~(\ref{eq:continuity_equation}). This strategy was followed, for
example, in Ref.\,\onlinecite{Reuter_Sondheimer_anomalous_Skineffect}.
Deng follows a similar strategy
[see end of \S\,3 in Ref.\,\onlinecite{Possible_Instability_Deng_2017}],
although he uses the continuity equation~(\ref{eq:wrong_formulation_continuity}) 
with the relaxation term.
Other, more recent examples of implementing conservation laws in the
Boltzmann equation can be found in Refs.\,\onlinecite{Roepke_1999,Atwal_2002}.
Atwal and Ashcroft, for example, compute hydrodynamic approximations to the
bulk plasmon dispersion relation.\cite{Atwal_2002}

The treatment of the near-surface charge density according to 
Eq.\,(\ref{eq:continuity-with-delta-source}) reveals another inconsistency.
It should contain two types of charges: 
surface charges with area density $\rho_{\rm s}$
that are (on the macroscopic scale) localised
at the surface $z = 0$, and a smooth charge density $\rho_{\rm b}(z)$ 
in the bulk $z > 0$ (see Sec.\,\ref{sec:review_BV_embed_Deng_response} for details). 
Indeed, the distribution $\Theta'(z)$ on the
rhs must have a surface charge as its pendant on the lhs; the 
divergence $\nabla \cdot \bm{J}$ being non-singular by 
Eq.\,(\ref{eq:complete_surface_description_half space_geometry}).
By splitting the continuity equation into bulk and surface
parts, one gets\cite{Bedeaux_Vlieger_opt_prop_surf}
\begin{eqnarray}
\partial_t \rho_{\rm b} + \nabla \cdot \bm{J}_{\rm b} &=& 0
\,,
\nonumber
\\
\partial_t \rho_{\rm s} + \nabla_\Vert \cdot \bm{J}_{\rm s} &=&
- J_{{\rm b},z}(\bm{x}_0)
\,,
\label{eq:bulk-surface-charge-conservation}
\end{eqnarray}
rather than Eq.\,(\ref{eq:continuity-with-delta-source}), with a surface
coordinate $\bm{x}_0$.
The localized current density $\bm{J}_{\rm s}$ is by consistency
parallel to the surface, hence only the parallel part $\nabla_\Vert$ of
the gradient. The second equation illustrates that a nonzero current 
$J_{{\rm b},z}(\bm{x}_0)$ signals a transfer from bulk to surface charge.
If we adopt a model with $\bm{J}_{\rm s} = \bm{0}$, the second line
of Eq.\,(\ref{eq:bulk-surface-charge-conservation}) yields 
Eq.\,(\ref{eq:continuity-charge-surface-current}) used above.

In the literature, this problem is handled
in different ways. In the local approximation, there is
only a surface charge which is generated by a jump in the Ohmic current 
$\sigma {\bm E}$. Models that treat the electron dynamics in more detail
lead in general to a non-local current-field relation [as 
in Eq.\,(\ref{eq:lin_resp_current})].
The majority of authors simply exclude any surface charge and use the
bulk component $\rho_{\rm b}$ in the first line of 
Eq.\,(\ref{eq:bulk-surface-charge-conservation})
to model a charge density localized within
the sub-surface region on some spatial scale that is actually resolved in 
the model. Typical candidates for this scale are the Thomas-Debye length
$v_{\rm F} / \omega_{\rm p}$ and the mean free path $v_{\rm F} \tau$. Consistency
with Eqs.\,(\ref{eq:bulk-surface-charge-conservation}) then requires 
the surface current $J_z(0)$ to vanish. Contrary to Deng's claims,
this does \emph{not} exclude the existence of a surface plasmon 
mode\cite{Ritchie_1963},
as we recapitulate in several examples throughout this paper. 

Alternative approaches keep both bulk and surface charge, but are then
in need of a model for the evolution of the surface charge that cannot
increase indefinitely.
We review recent examples in Sec.\,\ref{sec:From_HDM_to_LDM} within the 
framework of Eq.\,(\ref{eq:bulk-surface-charge-conservation}).
An explicit splitting of the charge into bulk and surface parts is not
manifest in Deng's papers. It appears in several places, however, that
he has in mind a charge sheet localized at the surface. The limiting
value $\rho_s = \lim_{q\to\infty} \rho_q$ of the cosine transform 
obviously provides its amplitude. If this is the only relevant charge,
the electric field behaves like ${\rm e}^{ - k z }$, but Deng notes
that this can only be used ``outside the layer of surface charges''
[after Eq.\,(30) and~(33) of Ref.\,\onlinecite{Possible_Instability_Deng_2017}].
{In Ref.\,\onlinecite{Deng_2019} he points out for a metal-dielectric 
interface where currents on both sides are different, that charges may
``pile up'' in the interface region (which is not resolved on the macroscopic
scale).} {Deng calls this
a ``capacitive effect'', having probably in mind the Coulomb energy in
this region of high charge density. Apart from that remark, no special
treatment is applied to secure that charge be conserved during the transfer
between bulk and surface. (In particular, a charge trap ensues if no desorption
from the interface region is implemented in the model.)}

Finally, Deng uses in the semiclassical model (SCM), based on the Boltzmann
equation, a boundary condition for the distribution function that combines
specular and diffuse scattering with probabilities $p$ and $1-p$, respectively.
We review this approach in the following section. 
It may suffice to say that the consistency of this
treatment with respect to charge conservation has been discussed in the
literature\cite{Zaremba_PhysRevB.9.1277, Keller_et_al_PhysRevB.12.2012, Cercignani_Lampis_Ofl_Streuung_RB} because it generates a nonzero surface current
that is excluded unless one allows for a genuine surface charge. 
Zaremba, for example, suggested that
the fraction $1-p$ of diffusely scattered electrons should not simply 
disappear from the current balance, but be described by a different velocity 
distribution.\cite{Zaremba_PhysRevB.9.1277} We mention in 
Sec.\,\ref{subsec:SCM__surface_problem} a general formulation in terms
of a boundary scattering kernel put forward in 
Ref.\,\onlinecite{Cercignani_Lampis_Ofl_Streuung_RB}.
The same criticism of an unphysical surface current has been formulated against 
the dielectric approximation
mentioned after Eq.\,(\ref{eq:translationally_variant_SRM_conductivity}),
see for example Ref.\,\onlinecite{Mead_1977}.

\section{\label{sec:SCM}Semiclassical (Boltzmann) model for electron dynamics}

A Boltzmann description for the response of metallic conduction electrons
has been used intensively in the past, starting with the anomalous
skin effect\cite{Reuter_Sondheimer_anomalous_Skineffect,Wagner_Oberfl_Wellen_e_Plasma}
where also a surface/interface problem had to be solved. 
We review in this section a solution to the Boltzmann equation 
at an interface, within the relaxation time approximation
and implementing local charge conservation. We recall in particular 
the handling of electron scattering at the metal surface.

In this section, we consider the case that the electronic density contains
only a ``bulk'' part in the sense of Eq.\,(\ref{eq:bulk-surface-charge-conservation}).
An alternative model system where volume electrons can be trapped
in the surface region and where a stationary state is defined 
by the balance between trapping of incoming electrons 
and desorption back into the bulk, 
will be treated in Sec.\,\ref{sec:From_HDM_to_LDM}.
Another approach to the surface plasmon dispersion relation that is 
not based on the Boltzmann equation, is presented in Sec.\,\ref{sec:pseudo_model}.

\subsection{The volume problem}
\label{subsec:SMC__volume_charge_conservation}

The Boltzmann equation determines the distribution function for the metal
electrons in phase space $f(\bm{v}, \bm{x}, t)$. The relaxation time
approximation replaces the collision term by
$( f_0 - f ) / \tau$ where $f_0$ is the equilibrium (Fermi-Dirac) distribution
that depends only on the electron energy.
Integrating over all velocities, one gets 
Eq.\,(\ref{eq:wrong_formulation_continuity}), understood as describing bulk charges
only, which does not locally conserve
the charge. Warren and Ferrell\cite{Warren_Ferrell_PhysRev.117.1252}
and
Mermin\cite{Mermin_PhysRevB.1.2362} showed that this can be 
repaired with a relaxation term
$( f_{\rm loc} - f) / \tau$ where $f_{\rm loc}$ differs from $f_0$ by a
shift in the Fermi energy such that the local electron density is
$n_0 + \delta n( \bm{x}, t )$. Using the scaling law for free electrons,
$\epsilon_\mathrm{F} \sim n_0^{2/3}$, one gets
\begin{equation}
f_{\mathrm{loc}}(\bm{v}, \bm{x}, t) 
= f_0 - f'_0 \dfrac{2\epsilon_{\mathrm{F}}}{3n_0} \delta n(\bm{x}, t)
\label{eq:local_thermodynamic_equilibrium_distribution}
\end{equation}
to first order in $\delta n$. Here $\epsilon_{\mathrm{F}}$ is 
the equilibrium Fermi energy and $f'_0$ the energy derivative of the 
Fermi-Dirac distribution.
Expanding the distribution function to first
order in the electric field, $f = f_0 + f_1 + \ldots$, the modified Boltzmann 
equation yields
\begin{align}
&
\left[\partial_t + \tau^{-1} + \bm{v}\cdot\nabla\right] f_1(\bm{v}, \bm{x}, t) 
\nonumber \\
& \qquad = - f'_0 \Big[e \bm{v} \cdot \bm{E}(\bm{x}, t)
+ \dfrac{2\epsilon_{\mathrm{F}}}{3n_0\tau}\delta n(\bm{x}, t)
  \Big] 
\,,
\label{eq:Energy_conv_Boltzmann_Mermin_equation} 
\end{align}
where $e$ is the electron charge. When both sides are integrated over velocity
space, we get
\begin{align}
\left[\partial_t + \tau^{-1}\right]\rho(\bm{x}, t) 
+ \nabla\cdot\bm{J}(\bm{x}, t) 
= \dfrac{e}{\tau}\delta n(\bm{x}, t) \,,
\end{align}
where the charge $\rho$ and current $\bm{J}$ densities are the zeroth and first 
moment of the perturbed velocity distribution $f_1$. If we now take $\rho = e \delta n$,
the terms involving $1/\tau$ cancel and local charge conservation follows:
\begin{equation}
\partial_t\rho(\bm{x}, t) + \nabla\cdot\bm{J}(\bm{x}, t) = 0 \,,
\label{eq:continuity_equation}
\end{equation}
rather than Eq.\,(\ref{eq:continuity-with-delta-source}). 
There is no localized source term here on the rhs (and the boundary
condition for the surface current is $J_z(\bm{x}_0) = 0$), 
unless one introduces
a surface component in the charge density, as in  
Eq.\,(\ref{eq:bulk-surface-charge-conservation}).

For the surface plasmon problem, the distribution function inherits 
the dependence $\exp {\rm i}(kx - \omega t)$ of the electric field,
and Eq.\,(\ref{eq:Energy_conv_Boltzmann_Mermin_equation}) yields
\begin{equation}
\left[ \partial_z + \eta \right] f_1(\bm{v}, z) 
= -\frac{ f'_0 }{ v_z } \Big[
e \bm{v} \cdot \bm{E}(z) 
+ \dfrac{2\epsilon_{\mathrm{F}}}{3 n_0 \tau} \delta n(z)
\Big]
\,. 
\label{eq:linear_Boltzmann_equation}
\end{equation}     
Here we have defined the (complex) inverse length 
$\eta = (1/\tau - {\rm i} \omega + {\rm i}  kv_x)/v_z$. 
To simplify the following calculations, we drop the
Mermin correction (the term with $\delta n$ on the rhs). It was also
not taken into account by Deng.

\subsection{\label{subsec:SCM__surface_problem}The surface problem}

We now discuss how the solutions to Eq.\,(\ref{eq:linear_Boltzmann_equation})
involve the boundary conditions at $z = 0$.
The general solution is
\begin{equation}
f_1(\bm{v}, z) = C(\bm{v})\,\mathrm{e}^{-\eta z} 
- \frac{f'_0}{v_z}\int_0^z dz'~e\bm{v}\cdot\bm{E}(z')~\mathrm{e}^{\eta(z'-z)} 
\,,
\label{eq:linear_Boltzmann_equation_solution}
\end{equation}
where the function $C(\bm{v})$ must be determined from the asymptotic
behaviour of $f_1$.
This is a standard 
procedure
\cite{Reuter_Sondheimer_anomalous_Skineffect,PhysRev.172.607,Wagner_Oberfl_Wellen_e_Plasma,Keller_et_al_PhysRevB.12.2012,Possible_Instability_Deng_2017}
that proceeds by considering separately the cases $v_z\lessgtr 0$.
For electrons
moving towards the surface, $v_z < 0$, causality requires that their
contribution can only depend on the electric field along their path in the
past, $z' > z$. Requiring that $f_1( \bm{v}, z )$ vanishes for $z \to \infty$,
we get
\begin{equation}
v_z < 0: \quad
C(\bm{v}) = \frac{f'_0}{v_z}\int_0^{\infty}dz'~e\bm{v}\cdot\bm{E}(z')
~\mathrm{e}^{\eta z'} \,.
\label{eq:Boltzmann_constant_integration_v_z_negative} 
\end{equation}
Deng
\cite{Possible_Instability_Deng_2017}
insists that the convergence of the integral in 
Eq.\,(\ref{eq:Boltzmann_constant_integration_v_z_negative}) is only
ensured when $\mathop{\mathrm{Re}} \eta < 0$. Allowing for a complex
frequency $\omega$, this yields for $v_z < 0$ the condition
$\mathop{\mathrm{Im}} \omega + 1/\tau > 0$. This does not imply an
instability ($\mathop{\mathrm{Im}} \omega > 0$), of course, 
and it is satisfied by standard calculations (see examples
in Sec.\,\ref{subsec:SCM_energy_conv_disp_rel} below).
\footnote{In the collisionless limit, $\tau \to \infty$, causality can be
enforced as before, by solving Eq.\,(\ref{eq:linear_Boltzmann_equation})
with the help electron trajectories arriving from the past at a given 
position. A complex frequency $\omega 
= \mathop{\rm Re}\omega + {\rm i} 0$
with an infinitesimal positive imaginary part does not signal an
instability neither, because it is just a convenient tool
to ensure an adiabatic switching-on of the electric field. The
amplitude of the latter is taken proportional to $\exp( \alpha t )$ 
with $\alpha \to +0$ so that the field vanishes in the remote past.}

The other case $v_z > 0$ includes electrons that are leaving the surface
and requires a model for surface scattering. The bulk version of the
solution~(\ref{eq:linear_Boltzmann_equation_solution}) would take in
that case $C(\bm{v}) = 0$ and shift the lower integration bound to 
$z' = -\infty$. Convergence at this lower limit would again be secured provided
$\mathop{\mathrm{Im}} \omega + 1/\tau > 0$. 
{For a medium bounded to $z \ge 0$, the integral term in
Eq.\,(\ref{eq:linear_Boltzmann_equation_solution}) describes 
electrons moving from $z'$ within the region $0 \le z' \le z$ 
to position $z$. The boundary term $C(\bm{v})$ 
takes into account electron trajectories that were moving towards the 
surface and are scattered back there.}
Deng uses the Fuchs parameter $p$
that gives the fraction of electrons undergoing specular 
reflection. \cite{fuchs_1938_conductivity_thin_metall_films,Reuter_Sondheimer_anomalous_Skineffect,PhysRev.172.607} 
This provides a way to model surface roughness 
which is inevitable for macroscopic samples.
The boundary condition formulated by Fuchs yields the simple result
\begin{equation}
v_z > 0: \quad
C(\bm{v}) = p \, C(\bm{v}_-)
\,,
\label{eq:historical_Fuchs_boundary_condition}
\end{equation} 
where $\bm{v}_- = \bfM\,\bm{v} = ( v_x, v_y, - v_z)$ is the mirror image of $\bm{v}$. 
Fuchs arrives at this ``$p$-model'' by assuming that the non-specularly reflected 
electrons are described by the isotropic equilibrium distribution 
function \cite{fuchs_1938_conductivity_thin_metall_films}.

The boundary condition formulated by 
Eq.\,(\ref{eq:historical_Fuchs_boundary_condition})
runs into the following problem, as pointed out in
Refs.\,\onlinecite{Zaremba_PhysRevB.9.1277, Cercignani_Lampis_Ofl_Streuung_RB}.
If we compute the surface current from 
Eqs.\,(\ref{eq:linear_Boltzmann_equation_solution}--\ref{eq:historical_Fuchs_boundary_condition}),
one gets a result equal to $(1-p)$ times 
the incoming current (the integral of $v_z f_1$ over the domain
$v_z < 0$ at $z = 0$) because the diffusively scattered electrons do
not appear in the distribution function. 
As pointed out in Sec.\,\ref{subsec:charge-conservation},
a nonzero surface current must be balanced by a charge sheet
located at the surface. In most applications of the Boltzmann equation,
it is assumed that the distribution function $f_1( \bm{v}, z )$
describes \emph{all} charges, and such a charge sheet is excluded. A
charge distribution of narrow, but finite extent arises within the
Boltzmann formalism on the length scale $\ell = v_\mathrm{F} \tau$ of the
mean free path where $v_\mathrm{F}$ is the Fermi velocity, but is included
in $f_1( \bm{v}, z )$. (An example can be seen in Fig.\,3(b) of
Ref.\,\onlinecite{Universal_Self_Amplification_Deng_et_al_2017}.)

These considerations have lead Zaremba
\cite{Zaremba_PhysRevB.9.1277} to correct the Fuchs $p$-model~(\ref{eq:historical_Fuchs_boundary_condition})  
of a partially diffuse boundary.
Motivated by this work and the general consideration of 
Cercignani and Lampis
\cite{Cercignani_Lampis_Ofl_Streuung_RB}, 
we now construct a $p$-model boundary condition that corresponds 
to a vanishing surface current.
The key concept is the probability
$S(\bm{v}' \mapsto \bm{v})$ for a transition 
from an incoming velocity $\bm{v}'$ to a
reflected (backscattered) one $\bm{v}$. One writes the balance
of outgoing and incoming currents
\begin{equation}
v_z > 0: \quad
v_z C(\bm{v}) = \int\limits_{v_z' < 0}\!{d}^3{v}'\,
(-v_z') C(\bm{v}') S(\bm{v}'\mapsto \bm{v}) 
\,.
\label{eq:CL_surface_scatt_scheme_particle_current}
\end{equation}
The condition $J_z(0) = 0$ is now ensured by the normalization integral
\begin{equation}
v_z' < 0: \quad
\int\limits_{v_z > 0} d^3v~S(\bm{v}'\mapsto\bm{v}) = 1
\,.
\label{eq:CL_surface_scatt_scheme_normalization}
\end{equation}   
A partially diffuse surface can be described by splitting
$S$ into a specularly reflecting part with weight $p$, proportional
to $\delta(\bm{v} - \bm{v}'_{-})$, and a diffuse part proportional to
$\cos\theta$ where $\theta$ is the angle of the final velocity $\bm{v}$
relative to the surface normal (the $z$-axis)
\cite{Zaremba_PhysRevB.9.1277, Keller_et_al_PhysRevB.12.2012}. By taking
the cosine to a higher power, the diffuse scattering would be preferentially
along the surface normal \cite{Zaremba_PhysRevB.9.1277}. In both cases,
there is no memory left of the direction of the incident velocity.
This {approximate scattering model} describes a surface which is rough on 
the scale of the
Fermi wavelength $\lambda_\mathrm{F}$ both in rms height and correlation length. 
Since $\lambda_\mathrm{F}$ is much smaller than the plasmon wavelength, 
the surface can still be viewed as smooth from a macroscopic perspective.
It is thus legitimate to assume translation
invariance so that $S$ does not depend on the in-plane coordinates $x, y$.
Assuming elastic scattering and 
keeping in mind the normalization~(\ref{eq:CL_surface_scatt_scheme_normalization}),
we arrive at
\begin{eqnarray}
&&
v'^2 S(\bm{v}'\mapsto\bm{v}) = \delta(v-v') \times
\nonumber\\
&& 
\qquad
\Big[ p ~ \delta(\phi - \phi')\delta(\cos\theta + \cos\theta')
    + \dfrac{1-p}{\pi}\cos\theta 
\Big]
\,,
\label{eq:normalized_p_model_scatt_prob}
\end{eqnarray}
where $\theta, \ldots \phi'$ are spherical coordinates relative to the surface normal.
Inserting this into Eq.\,(\ref{eq:CL_surface_scatt_scheme_particle_current}),
we obtain the boundary condition
\begin{equation}
v_z > 0: \quad
C(\bm{v}) = p \, C(\bm{v}_-) + (1-p) A(v)
\label{eq:advanced_surface_scatt_boundary_cond}
\end{equation}
with
\begin{equation}
A(v) = \frac{1}{\pi} \int\limits_{\cos\theta'<0}\!d\Omega'\,
(-\cos\theta') C(v,\theta',\phi') 
\,,
\label{eq:advanced_surface_scatt_boundary_cond_A}
\end{equation}
where $\mathrm{d}\Omega' = \sin\theta'\mathrm{d}\theta' \mathrm{d}\phi'$
and $C$ is given by 
Eq.\,(\ref{eq:Boltzmann_constant_integration_v_z_negative}).
This is a simple additional boundary condition (ABC) to be used in conjunction
with the Boltzmann equation when there is no accumulation of charges at the
surface.
The additional term $(1-p) A(v)$ in 
Eq.\,(\ref{eq:advanced_surface_scatt_boundary_cond}) 
describes the diffuse scattering; it takes care of balancing the currents
of incoming and outgoing charges.\cite{Cercignani_Lampis_Ofl_Streuung_RB}
The procedure leading to this ABC illustrates also that the response of a surface 
is a problem on its own: it cannot be solved from information about the
bulk behaviour alone. It is only conceivable within microscopic models that
no ABC is needed: in that case, the behaviour of the electronic wave functions
in the surface region embodies the (additional) boundary condition. This would
be the viewpoint of density functional theory,\cite{Liebsch_1997} for example.

One may ask why
Deng did not run into contradictions in the specular case $p = 1$ 
where Eqs.\,(\ref{eq:historical_Fuchs_boundary_condition},\ref{eq:advanced_surface_scatt_boundary_cond}) coincide and lead to
$J_z(0) = 0$ (by mirror symmetry with respect to the plane $z = 0$). 
He comments on this specular reflection model
in Appendix~B of Ref.\,\onlinecite{Deng_2019} where
the nature of the fictitious charge sheet is analyzed, which we have
alluded to after Eq.\,(\ref{eq:SRM_mirror_symmetric_electric_field}).
The existence of surface plasmons is connected to a divergence
of the fictitious charge amplitude.
Deng manages to transform this condition into the dispersion relation 
of Eq.\,(\ref{eq:Deng_def_surface_diel_fctn}) 
by suitably identifying the bulk
dielectric function, although the division by $J_z(0) = 0$
is illegitimate.

\subsection{Approximate solution}
\label{subsec:Boltzmann-approximate-solution}

We now determine the actual distribution function for a surface plasmon problem. 
The equations determined so far are still lacking a specific form for the
electric field. We compute its potential $\phi$ by solving
Poisson's equation $\bm{\Delta} \phi = -4\pi \rho$. Since we are interested in a solution with zero surface current, only volume charge is present. Keeping in mind that in our geometry 
$\nabla = ({\rm i} k, 0, \partial_z)^{\sf T}$ and using the method of Green's functions, 
we arrive at
\begin{align}
\phi(z) = \dfrac{2\pi}{k}\int_0^{\infty}\!d z'~ \mathrm{e}^{-k|z-z'|} \rho(z')
\label{eq:electrostatic_field_Greens_function}
\end{align}
within the metal half space.
Rather than the cosine transform used by Deng, we use the Fourier expansion
\begin{equation}
\rho(z) = \int_{-\infty}^{\infty}\dfrac{d q}{2\pi} 
\,\mathrm{e}^{{\rm i} qz} \tilde\rho(q)
\label{eq:charge-in-Fourier-expansion}
\end{equation}
for the charge density $\rho$. It turns out that the charge density varies
on length scales $v_{\rm F}/\omega_{\rm p}$ much shorter than the plasmon
wavelength $1/k$. Therefore, Deng introduces the approximation of a 
charge sheet insofar as it appears
under the integral~(\ref{eq:electrostatic_field_Greens_function}).
Using the form $\rho(z)\approx \rho_{\rm s}\delta(z - d)$ with $d > 0$
slightly located in the metal,
we obtain the fields
\begin{eqnarray}
\phi(z) &=&
\frac{ 2\pi \rho_{\rm s} }{ k } \,
{\rm e}^{-k|z - d|}
\label{eq:electric_potential_displaced_charge}
\\
\left(\begin{matrix} E_x(z) \\ E_z(z) \end{matrix}\right) 
&=& 
-2 \mathrm{i} \rho_{\rm s} \int dq~\left(\begin{matrix} k \\ q \end{matrix}\right) \dfrac{\mathrm{e}^{{\rm i} q(z-d)}}{k^2+q^2} 
\nonumber\\
&=& 2 \pi \rho_{\rm s} \left(\begin{matrix} -{\rm i} \\ \mathop{\mathrm{sgn}}(z - d) \end{matrix}\right) 
\mathrm{e}^{ - k |z - d| }
\,,
\label{eq:electric_field_displaced_charge}
\end{eqnarray}
where 
$\mathop{\mathrm{sgn}}$ is the sign function.
The $z$-component of this expression is discontinuous at $z = d$. By allowing
for $d > 0$, 
we can control carefully how to interchange the $q$-integral 
with the $z$-integral of Eq.\,(\ref{eq:linear_Boltzmann_equation_solution}):
according to the residue theorem, the position relative to the charge 
layer at $z = d$ decides where to close the integration path in the complex $q$-plane. 
Deng takes $d = 0$ and seems to ignore this 
difficulty \cite{Possible_Instability_Deng_2017}.

The following calculations are straightforward: 
we put the electric field~(\ref{eq:electric_field_displaced_charge})
into the solution~(\ref{eq:linear_Boltzmann_equation_solution}) of
the Boltzmann equation, use the boundary 
conditions~(\ref{eq:advanced_surface_scatt_boundary_cond}) 
and~(\ref{eq:advanced_surface_scatt_boundary_cond_A})
and integrate separately over the surface region 
($0<z<d$) and the bulk of the metal ($z>d$). Eventually we take the 
limit $d \to 0$. We use this scheme 
in Sec.\,\ref{sec:Calculation_amplification_loss_rate}
to compute the power loss of the surface plasmon. One finds that
the contributions from the region $0<z<d$ vanish when the limit $d \to 0$
is taken. We therefore present in the following the results in this
limit only.

\begin{table*}[t!h]
\begin{tabular}{ll|l|l}
\hline
Type & 		& ``$F_\pm$'' 		& ``$F_0$''
\\[0.5ex]
\toprule
``bulk'' & $g_b$\,\,		
	& $\displaystyle
	- 2 \,{\rm e}^{ -k z} 
	F_0( \bm{k}, \bar{\omega}, \bm{v} )
	$
	& $\displaystyle
	+ {\rm e}^{ - k z } F_0( \bm{k}, \bar{\omega}, \bm{v} )
	$
\\
	& & $\displaystyle {}	
	+ 2 \,\Theta(v_z) 
	\,{\rm e}^{ - \eta z} 
	\left[
	F_0( \bm{k}, \bar{\omega}, \bm{v} )
	- F_0( \bm{k}, \bar{\omega}, \bm{v}_- )
	\right]
$		
	& 
\\[0.5ex]
\hline
``surface'' & $g_s$	\,\,	
	& $\displaystyle
+ 2 \,\Theta( v_z) \,{\rm e}^{ -\eta z } (1 - p) 
F_0( \bm{k}, \bar{\omega}, \bm{v}_- )
$		
	& $\displaystyle
-\Theta( v_z) \,{\rm e}^{ -\eta z }
\left[
F_0( \bm{k}, \bar{\omega}, \bm{v} )
- p F_0( \bm{k}, \bar{\omega}, \bm{v}_- )
\right]
	$
\\[0.5ex]
\hline
total &
	& \multicolumn{2}{l}{$\displaystyle\hspace*{20ex}
	- {\rm e}^{ -k z} F_0( \bm{k}, \bar{\omega}, \bm{v} )
	$}
\\
 & & \multicolumn{2}{l}{$\displaystyle\hspace*{20ex}
	{} + \Theta( v_z) \,{\rm e}^{ -\eta z }
	\left[
	F_0( \bm{k}, \bar{\omega}, \bm{v} )
	- p F_0( \bm{k}, \bar{\omega}, \bm{v}_- )
	\right]
	$}
\\
\hline
\end{tabular}
\caption[]{%
Contributions to the electronic distribution function
according to the labels used in Refs.\,\onlinecite{Possible_Instability_Deng_2017,Deng_2019}. 
The {table rows}
correspond to Eqs.\,(\ref{eq:Deng-g-bulk}, \ref{eq:Deng-g-surface})
in Appendix~\ref{a:check-Deng-gsgb},
the columns to the terms involving the functions
$F_0$ and $F_\pm$ defined in 
Eq.\,(\ref{eq:def-F0})
and Eqs.\,(\ref{eq:def-Deng-F0}, \ref{eq:def-Deng-Fplus-minus}).
We collect
the results of the ${\rm d}q$-integrations using the approximation
$\rho_q \approx \rho_{\rm s}$. 
Evaluating the integrals with the residue theorem, the ``$F_\pm$ terms'' 
turn into the ``$F_0$ type.'' 
A common factor $2\pi e \rho_{\rm s} f'_0 / k$ is taken out.
}
\centerline{\rule{\textwidth}{0.5pt}}
\label{t:Deng-gb-gs-terms}
\end{table*}

The procedure generates the following result for the boundary term 
$C(\bm{v}$) [Eq.\,(\ref{eq:Boltzmann_constant_integration_v_z_negative})]:
\begin{align}
v_z < 0: \quad
C(\bm{v}) & = - \frac{ 2\pi e \rho_{\rm s} f'_0 }{ k }
F_0( \bm{k}, \bar{\omega}, \bm{v} )
\label{eq:result-C-of-v}
\\
F_0( \bm{k}, \bar{\omega}, \bm{v} ) &=
\frac{ \bm{k} \cdot \bm{v} }{\bar{\omega} - \bm{k} \cdot \bm{v} }
\,,
\label{eq:def-F0}
\end{align}
where $\bm{k} = (k, 0, {\rm i}k )$ and $\bar{\omega} = \omega + {\rm i}/\tau$.
The distribution function itself comes as a formidable sum of three terms 
as follows
\begin{subequations}
\label{eq:metal_volume_distribution_fctns}
\begin{eqnarray}
\dfrac{
f_{\mathrm{b}}(\bm{v}, z) }{
2\pi e \rho_{\rm s} f'_0
}
&=& - \frac{ \mathrm{e}^{-kz} }{ k }
F_0( \bm{k}, \bar{\omega}, \bm{v} )
\,,
\label{eq:SCM_distribution_f_b}
\\
\dfrac{
f_{\mathrm{s,F}}(\bm{v}, z) }{
2\pi e \rho_{\rm s} f'_0
}
&=& \Theta(v_z)
\, \frac{ \mathrm{e}^{-\eta z} }{ k }
\Big[
F_0( \bm{k}, \bar{\omega}, \bm{v} )
\nonumber\\
&& {} \qquad 
- p\, 
F_0( \bm{k}, \bar{\omega}, \bm{v}_- )
\Big]
\,,
\label{eq:SCM_distribution_f_s}
\\
\dfrac{
f_{\mathrm{s,Z}}(\bm{v}, z) }{
2\pi e \rho_{\rm s} f'_0
} &\approx& 
\Theta(v_z) \, \mathrm{e}^{-\eta z }
(1 - p) \dfrac{2 v}{\bar{\omega} }  
\bigg(
\dfrac{ {\rm i} }{ 3 } + 
\dfrac{ k v }{ 8\, \bar{\omega} }
\bigg)
   \,,
\label{eq:SCM_distribution_f_A}
\end{eqnarray}
\end{subequations}
where 
$\eta = -{\rm i}(\bar{\omega} - k v_x)/v_z$, as defined after
Eq.\,(\ref{eq:linear_Boltzmann_equation}).
Here, $f_{\mathrm{b}}$ represents the volume solution in an infinitely extended metal
to a field with momentum $\bm{k}$,
$f_{\rm s,F}$ describes the specularly scattered fraction in the 
Fuchs model [Eq.\,(\ref{eq:historical_Fuchs_boundary_condition})], 
and $f_{\mathrm{s,Z}}$ takes into account diffuse scattering within Zaremba's
model [second term of Eq.\,(\ref{eq:advanced_surface_scatt_boundary_cond})].
In evaluating this last term, we have taken the two leading terms for small
$kv_{\rm F}/\bar{\omega}$.

\subsection{Comparison to Deng}
\label{s:check-Deng-gsgb}

Expressions of similar complexity are also found in 
Refs.\,\onlinecite{Possible_Instability_Deng_2017,2018arXiv180608308D,Deng_2019},
although a Fourier/cosine integral over $\rho_q$ 
is performed in the last step.
In concrete evaluations, the same simplification as here is applied, 
see for example from Ref.\,\onlinecite{Possible_Instability_Deng_2017}:
``using $E_z(z) \approx 2 \pi \rho_\mathrm{s} \mathrm{e}^{-kz}$ 
outside the layer of surface charges'' [i.e. for $z > d \to 0$ in our notation,
see Eq.\,(\ref{eq:electric_field_displaced_charge})].
Our calculation takes advantage of evaluating this integral
with the residue theorem first
and thus avoids convergence problems at large $q$. 
To ensure convergence at large $q$, Deng indeed has to introduce 
a cutoff $q_c \sim \omega_{\rm p} / v_{\rm F}$.

From the boundary conditions to the Boltzmann equation stated above,
we may thus expect that the terms $f_{\mathrm{b}} + f_{\mathrm{s,F}}$ 
correspond essentially with Deng's expressions, while $f_{\mathrm{s,Z}}$
is a correction that takes care of charge conservation at the partially diffuse 
surface. In Appendix~\ref{a:check-Deng-gsgb}, we check that this is indeed 
the case, and the results, according to Deng's notation, are displayed
in Table~\ref{t:Deng-gb-gs-terms}. It is remarkable to which extent the 
splitting into ``bulk'' and ``surface'' is ambiguous. 
In the following section, we evaluate the amplification (or loss)
rate of the surface plasmon using the energy balance argument
developed in Ref.\,\onlinecite{Possible_Instability_Deng_2017}.

\section{\label{sec:Calculation_amplification_loss_rate}Surface Plasmon Amplification Rate}

\subsection{Electric energy balance}
\label{subsec:energy-balance}

The amplification rate of the surface plasmon is determined 
by Deng\cite{Possible_Instability_Deng_2017} 
by a power balance equation that is derived from charge conservation.
Eq.\,(\ref{eq:continuity-with-delta-source}) is multiplied by
$\phi(\bm{x},t)$ and integrated over $\bm{x}$. 
We find two differences with respect to Deng's argument. First, the
collision term $\rho/\tau$ on the lhs should be suppressed, as charge must
be conserved locally. Second, when the ``$\nabla \cdot \bm{J}$ term''
is integrated by parts, we get
\begin{equation}
\int\limits_{z\ge0}\!d^3x \, \phi ( \nabla \cdot \bm{J} ) = 
- \int\!dA \, \phi( \bm{x}_0 ) J_z( \bm{x}_0 )
+
\int\limits_{z\ge0}\!d^3x \, \bm{E} \cdot \bm{J} 
\,.
\label{eq:partial-integration}
\end{equation}
The first term on the rhs is a surface integral at $z = 0$ (with
coordinates $\bm{x}_0$), while the other boundary, deep in the bulk, 
does not contribute, of course. This boundary term 
cancels exactly the integral over the singular term 
$- \Theta'(z) J_z( \bm{x}_0 )$ on the
rhs of Eq.\,(\ref{eq:continuity-with-delta-source}):
\begin{equation}
- \int\!d^3x \,\Theta'(z) \phi(\bm{x}) J_z( \bm{x}_0 )
=
- \int\!dA \, \phi( \bm{x}_0 ) J_z( \bm{x}_0 )
\,.
\label{eq:surface-integral}
\end{equation}
We emphasize that the surface current $J_z( \bm{x}_0)$ 
drops out at this stage from the energy balance which takes the form
\begin{equation}
\int\limits_{z\ge0}\!d^3x \, \phi \partial_t \rho
+
\int\limits_{z\ge0}\!d^3x \, \bm{E} \cdot \bm{J} 
= 0
\,.
\label{eq:}
\end{equation}
To proceed, we use the exponential $x$- and $t$-dependence of all fields
\footnote{The physical fields are taken as the real part of the
complex ones. This explains the additional factor $1/2$ in 
Eq.\,(\ref{eq:energy_power_densities}).}
to re-write the term $\phi \partial_t \rho$ as the derivative of an
electrostatic energy. 
Factoring out the surface area, one gets the balance equation
\(
\partial_t {\mathcal{E}} + {\mathcal{P}} = 0 ,
\)
where ${\mathcal{E}}$ and ${\mathcal{P}}$ are the electrostatic potential
energy and power per unit area. 
Decomposing the frequency into real and imaginary parts, 
$\omega = \omega_{\rm s} + {\rm i} \gamma$, we get
\begin{equation}
\gamma = - \frac{ \mathcal{P} }{ 2\mathcal{E} }
\label{eq:energy_conversion_imaginary_part_frequency}
\end{equation}
with energy and power per unit area, averaged over one oscillation cycle, given by 
\begin{subequations}
\label{eq:energy_power_densities}
\begin{eqnarray}
\mathcal{E} &=& 
\frac{1}{4} \int dz~ 
\mathop{\mathrm{Re}}\big[ \phi^*(z) \rho(z) \big] \, {\rm e}^{2 \gamma t}\,,
\label{subeq:electrostatic_energy_definition}
\\
\mathcal{P} &=& 
\frac{1}{2} \int dz~ 
\mathop{\mathrm{Re}}\big[ \bm{E}^*(z) \cdot \bm{J}(z) \big]\, {\rm e}^{2 \gamma t}
\,.
\label{subeq:electrostatic_power_definition}
\end{eqnarray}
\end{subequations}
A positive $\gamma$ would signal an instability where energy from
the electronic system is converted into a surface plasmon oscillation.

If we compare Eq.\,(\ref{eq:energy_conversion_imaginary_part_frequency}) 
with Eq.\,(8) of Ref.\,\onlinecite{Possible_Instability_Deng_2017}, 
we see that on the lhs, just $\gamma$ rather than $\gamma + 1/\tau$ appears,
which is due to the collision term in question. Deng adds on the
rhs the power (per area) due to the surface current
\begin{equation}
{\mathcal{P}}_{\mathrm{sc}} 
= \frac{1}{2}\mathop{\mathrm{Re}}\big[ \phi^*(0) J_z(0) \big] \, {\rm e}^{2 \gamma t}
\label{eq:surface-power-Ps-definition}
\end{equation}
which should, as mentioned above, 
cancel with the boundary term from the partial integration 
[Eq.\,(\ref{eq:partial-integration})].
As the surface power ${\mathcal{P}}_{\rm sc}$ plays a key role in 
the claimed surface plasmon instability of
Ref.\,\onlinecite{Possible_Instability_Deng_2017}, 
this cancellation is a crucial point. We discuss its contribution
in Sec.\,\ref{sec:contribution_to_power}, 
where we add up the imaginary parts of 
the frequency. 

The energy balance condition~(\ref{eq:energy_conversion_imaginary_part_frequency})
provides a simple way to picture a potential instability of the surface plasmon.
With $\gamma > 0$ and from Eq.\,(\ref{subeq:electrostatic_power_definition}),
it would feature 
a current density opposite to the electric field (and in phase with it)
because $\mathcal{E}$ will always be positive. 
Now, due to partial reflection at the inner surface, it is possible that
a fraction of charges indeed flows ``against the field''. 
Ritchie and Marusak phrase their analysis into the following, however:
``The origin of surface plasmon damping in the present approximation
[treating the electrons within the Boltzmann equation] lies in the fact
that the surface, which is assumed infinitely massive, is able to absorb
momentum. Thus an electron in the semi-infinite gas may collide with the surface,
lose momentum to it, and then may be able to interact with a surface
plasmon.'' [Ref.\,\onlinecite[end of \S\,1]{Ritchie_Marusak}]
We detail
in the following the different contributions to the power $\mathcal{P}$.

\subsection{\label{sec:contribution_to_power}Discussion of the amplification rate}
\label{subsec:SCM_energy_conv_disp_rel}

From the potential~(\ref{eq:electric_potential_displaced_charge})
we get for the energy density~(\ref{subeq:electrostatic_energy_definition}):
\footnote{The integrals runs over the entire charge density which may
be thought as being placed in a region $0 < z < d$, see discussion 
after Eq.\,(\ref{eq:charge-in-Fourier-expansion}). Deng defines the
energy and power densities 
by extracting a factor $1/2$, see Eq.\,(3) in Ref.\,\onlinecite{Possible_Instability_Deng_2017}.
}
\begin{equation}
\mathcal{E} = \frac{ \pi |\rho_{\rm s}|^2 }{ 2 k }\, {\rm e}^{2 \gamma t}
\,.
\end{equation}
The contributions to the distribution function listed in 
Eq.\,(\ref{eq:metal_volume_distribution_fctns}) 
generate the current density $\bm{J}$ needed for the power $\mathcal{P}$
[Eq.\,(\ref{subeq:electrostatic_power_definition})]. 
To simplify the velocity integrals, 
we continue to expand consistently 
to the first order 
in $k v_{\rm F} / \omega_{\rm s}$ and
$\gamma/\omega_{\rm s}$ and take for the real part
$\omega_{\rm s} = \omega_{\rm p} / \sqrt{2}$,
the well-known long-wavelength limit of the surface plasmon 
frequency.\cite{Feibelman_1971,Flores_1972}

Referring to the three parts of the distribution function
[Eq.\,(\ref{eq:metal_volume_distribution_fctns})], we get a self-consistent
equation with three terms
\begin{equation}
\gamma = \gamma_{\mathrm{b}} + \gamma_{\mathrm{s, F}} + \gamma_{\mathrm{s, Z}}
\,,
\label{eq:imag_freq_Fuchs_model}
\end{equation}namely
\begin{subequations}\label{eq:imag_freq_Fuchs_model_contributions}\begin{eqnarray}
	\gamma_{\mathrm{b}} &=& - \gamma - \dfrac{1}{\tau} \,,
	\label{subeq:imag_freq_volume} \\
	\gamma_{\mathrm{s,F}} &=& 
	 - \dfrac{3}{4}kv_{\mathrm{F}}
	 + \dfrac{3(1 - p)}{16}kv_{\mathrm{F}} \,,
	\label{subeq:imag_freq_surface_Fuchs} 
	\\
	\gamma_{\mathrm{s,Z}} &=& - \dfrac{1-p}{3}kv_{\mathrm{F}} \,.
	\label{subeq:imag_freq_surface_Zaremba} 
\end{eqnarray}
\end{subequations}
The bulk contribution $\gamma_{\rm b}$ follows from the 
factor $1/\bar{\omega}$ in the distribution 
function~(\ref{eq:metal_volume_distribution_fctns}).
When inserted into Eq.\,(\ref{eq:imag_freq_Fuchs_model}), it effectively halves the
amplification rate $\gamma$.
All contributions provide a damping of the surface plasmon, 
so that a detailed comparison to Deng's results \cite{Possible_Instability_Deng_2017}
is in order.
This requires some care, since our calculation is organised in a different 
manner. In particular, the assignment of terms to
``bulk'' and ``surface'' is not unique (see Table~\ref{t:Deng-gb-gs-terms}). 
Recall, for example, the first line
in Eq.\,(\ref{eq:translationally_variant_SRM_conductivity}) which resembles
a bulk contribution, although it acquires a surface character because the
integral is cut off at the metal surface (dielectric approximation). 
We therefore aim at discussing the complete result(s) for the amplification
rate. Its dependence on the plasmon momentum $k$ may also guide our physical 
insight.

First of all, the terms depending on the relaxation rate $1/\tau$ 
cannot be compared in a meaningful way because in 
Ref.\,\onlinecite{Possible_Instability_Deng_2017}, the equation of 
continuity in the bulk system [Eq.\,(\ref{eq:wrong_formulation_continuity})] 
contains an unphysical charge relaxation. 

Second, the behaviour of the current at the surface is completely different.
As mentioned after Eq.\,(\ref{eq:partial-integration}), the contribution 
from the jump in the current density, $\mathcal{P}_{\rm sc}$ 
[Eq.\,(\ref{eq:surface-power-Ps-definition})],
drops out from the energy balance.
In Deng's papers, however, it plays a key role
[see Ref.\,\onlinecite[Sec.\,6]{Possible_Instability_Deng_2017}]. 
He finds that the contributions to lowest order in $k v_{\rm F} / \omega_{\rm s}$ 
cancel each other in the sum $\mathcal{P} + \mathcal{P}_{\rm sc}$. This
is equivalent to removing the term $\gamma_{\mathrm{b}}$ in 
Eq.\,(\ref{eq:imag_freq_Fuchs_model}) and leads to amplification rates
that come out twice as large as in our calculation. 

Equipped with this
rule, we can analyze the contribution to the amplification rate that
Deng would reach from the power $\mathcal{P}$, had he ignored the surface current.
From the last equation (not numbered) in 
Ref.\,\onlinecite[Sec.\,5]{Possible_Instability_Deng_2017} (with $p = 1$):
\begin{equation}
\text{Deng, $\mathcal{P}$ only:}\quad
\gamma = - \frac{ 1 }{ \tau } - \frac{3}{2} k v_{\rm F}
\,.
\label{eq:}
\end{equation}
The last term is twice as large as the first term of $\gamma_{\rm s,F}$, but
has the same sign (plasmon damping). It is interpreted by Deng as the
contribution of Landau damping because it arises from a pole in his $q$-integrals
at $q = {\rm i} \eta$. (This condition
is indeed equivalent to $\bar{\omega} = k v_x + q v_z$ 
where an electron moves in phase with the electric field.)
This is consistent with our calculation,
since the same term appears from a part of the distribution 
function that varies with ${\rm e}^{ -\eta z}$ [$f_{\rm s,F}$, see 
Eq.\,(\ref{eq:SCM_distribution_f_s})]. 

It is interesting to note that the term proportional to $1-p$ in 
Eq.\,(\ref{subeq:imag_freq_surface_Fuchs}) is positive.
Since it is related to the fraction of charges that are not specularly reflected,
it seems to confirm Deng's picture that the symmetry breaking by the
surface is essential for the amplification of the plasmon mode. 
This does not hold in our calculation, however, because one has to add
the term~(\ref{subeq:imag_freq_surface_Zaremba}) to avoid having a charge
sink at the surface. The sum of the two contributions is negative,
$3/16 - 1/3 
= -7/48$, so that also the diffuse scattering provides an overall damping
channel.

Third, let us try to get an idea what would change when the surface
current contribution $\mathcal{P}_{\rm sc}$ to the power had been kept.
This may give a semi-quantitative estimate of the error incurred in
Deng's calculation. In our calculation, the surface model is
constructed in such a way that the surface current $J_z(0) = 0$
(see discussion in Sec.\,\ref{subsec:SCM__surface_problem}). It appears 
by inspection that Deng's solution for the distribution function
only contains the terms called $f_{\rm b}$ and $f_{\rm s,F}$
listed in Eq.\,(\ref{eq:metal_volume_distribution_fctns}): he uses
indeed the Fuchs boundary condition~(\ref{eq:historical_Fuchs_boundary_condition})
without further modifications. This means that we can use the
correction term $f_{\rm s,Z}$ to estimate Deng's result for the 
surface current (note the minus sign)
\begin{equation}
\text{Deng:}\quad
J_z(0) = - N \int\!d^3v\, e v_z f_{\rm s,Z}( \bm{v}, z = 0 )
\,.
\label{eq:guessing-Dengs-surface-current}
\end{equation}
Here, $N = (m / 2\pi\hbar)^3$ is a scale factor for the velocity
integral when the distribution function $f_0$ is taken as the Fermi-Dirac
distribution.
This calculation is fairly easy because 
the velocity dependence of $f_{\rm s,Z}( \bm{v}, z = 0 )$ is simple
[see Eq.\,(\ref{eq:SCM_distribution_f_A})], and we get
\begin{eqnarray}
\text{Deng:}\quad
J_z(0) &=&
\frac{ 1 - p }{ 4 } 
\omega_{\rm p}^2 \rho_{\rm s}
\bigg(
\frac{ \mathrm{i} }{ \bar{\omega} } + 
\dfrac{ 3 k v_F }{ 8\, \bar{\omega}^2 }
\bigg)
\,.
\label{eq:guess-Deng-Jz}
\end{eqnarray}
The scaling with the diffuse scattering fraction $1-p$ is as expected
from the qualitative discussion in 
Sec.\,\ref{subsec:SCM__surface_problem}: it corresponds to the `missing
charge' that is not (specularly) reflected.
The corresponding power gets contributions from the $1/\bar{\omega}$
term and a term linear in $k v_F$. Working out the ratio to the 
electrostatic energy, we find
\begin{equation}
\text{Deng:}\quad
\gamma_{\rm sc} 
= 
- \frac{ 1 - p }{ 4 }
\bigg( \gamma + \frac{ 1}{\tau} + \dfrac{ 3 k v_F }{ 8 } \bigg)
\,.
\label{eq:result-gamma-sc}
\end{equation}
This is also a damping contribution, in distinction to the claim
around Eq.\,(42) from Ref.\,\onlinecite{Possible_Instability_Deng_2017}.
If we added it to the three terms in Eq.\,(\ref{eq:imag_freq_Fuchs_model}),
we would get
a similar structure as in Deng's Eq.\,(44) where the amplification rate
$\gamma$ must be computed in a self-consistent way.

\begin{figure}[b]
\centerline{\includegraphics*[width = 0.83\columnwidth]{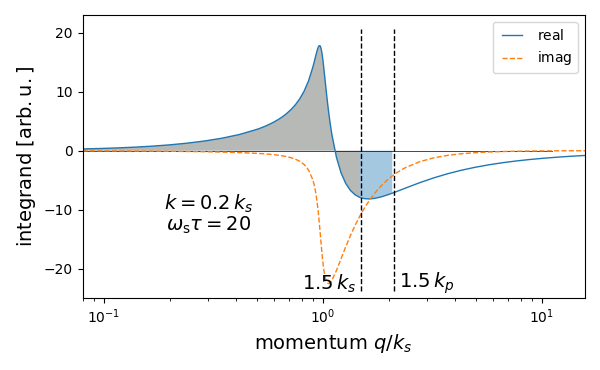}}
\vspace*{-2ex}
\caption[]{Integrand of Eq.\,(\ref{eq:Dengs-Gamma}) 
for the integration over
$q$ (we added the contributions from $q$ and $-q$, improving the
UV convergence). The singularity appears at $K v_F = \omega_{\rm s}$ 
with $K = \sqrt{ k^2 + q^2 }$, but it is smoothed by the imaginary part 
of $\bar{\omega}$. 
The vertical lines illustrate UV cutoffs quoted
in Refs.\,\onlinecite{Universal_Self_Amplification_Deng_et_al_2017,
Deng_2019}. We assumed a ratio $\rho_q / \rho_s \approx 1$,
the momentum scale is $k_s = \omega_{\rm s} / v_F$.}
\label{fig:sketch-integrand-42}
\end{figure}

Let us finally illustrate the poor convergence at large $q$ of Deng's
amplification rate, Eq.\,(42) from Ref.\,\onlinecite{Possible_Instability_Deng_2017}.
The term that is claimed to ``dominate all the contributions from other parts 
of $\mathcal{P}$ and $\mathcal{P}_{\rm sc}$'' is given by 
[the
scale factor $N$ was defined after Eq.\,(\ref{eq:guessing-Dengs-surface-current})]
\begin{eqnarray}
\Gamma &=& \frac{ 1-p }{ \rho_{\rm s} }
\int\limits_{-\infty}^{\infty}\!\frac{ dq\, 4 \rho_q }{ q^2 + k^2 }
\times
\nonumber
\\
&& {} 
N \int\limits_{v_z > 0}\!d^3v ( -e^2 f'_0 )
v_z \mathop{\rm Re}
\frac{ \bm{K} \cdot \bm{v} }{ \bar{\omega} }
\frac{ \bm{K} \cdot \bm{v} }{ \bar{\omega} - \bm{K} \cdot \bm{v} }
\,,
\label{eq:Dengs-Gamma}
\end{eqnarray}
where $\bm{K} = ( k, 0, q )$ and $\rho_{q}$ is extended in an even way
to $q < 0$.
From the calculations displayed by Deng, it is clear that this term
arises by subtracting its limit for small $K v_F / \omega_{\rm s}$.
This leads, however,
to a poor convergence of the $q$-integral, as can be easily seen by performing
the $\bm{v}$-integration first. 
The result of this (numerical) calculation is shown in 
Fig.\,\ref{fig:sketch-integrand-42}: the Landau peak at 
$q \sim k_s = \omega_s/v_F$ is close to the cutoff momentum $q_c$ 
[the value $q_c = 1.5\,k_s$ is taken from Ref.\,\onlinecite{Universal_Self_Amplification_Deng_et_al_2017}]. From
the calculation of integrals like~(\ref{eq:Dengs-Gamma})
with the residue theorem [see Appendix~\ref{a:check-gb-gs}],
we expect that the result is mainly imaginary [up to small corrections
${\cal O}( 1/\omega_s \tau)$]. Consistent with this is the visual
impression in Fig.\,\ref{fig:sketch-integrand-42}
that the areas under the real part of the integrand cancel. 

This can be checked by performing 
the $dq$-integral first, leading to the
entry ``$g_b \mid F_\pm$'' in Table~\ref{t:Deng-gb-gs-terms}, and then
evaluating the $d^3v$-integral. We find that this term contributes to the damping
rate the expression given in Eq.\,(\ref{eq:result-gamma-sc}), multiplied by
$-2$. The correction to the local approximation is of
relative order $k v_F / \omega_s$, in contradiction to the claim in
Ref.\,\onlinecite[Eq.\,(43)]{Possible_Instability_Deng_2017}.
Taken alone, this term would be interpreted as amplification. 
It is, however, just one contribution. 
There are terms that are dropped in Deng's
calculation of the ``surface power'' ${\cal P}_{\rm sc}$, for example
those that appear proportional to $\sin q z$ in an integral representation of 
the electric field 
[Eq.(\ref{eq:Deng-Ez-dq-integral})].
Although this term appears to vanish at $z = 0$, the $q$-integral may actually 
generate a function that is discontinuous at $z = 0$ and whose limiting value
for $z \downarrow 0$ is nonzero (see the Appendix~\ref{as:cos-and-exp-integrands}).
This discussion is somewhat futile, since,
recalling the argument of Sec.\,\ref{subsec:energy-balance}, 
the ``surface power'' ${\cal P}_{\rm sc}$ should not be counted 
at all in the energy balance.

To summarize this section: we have performed an estimation of the surface plasmon
amplification rate along an energy balance scheme put forward by
Deng.\cite{Possible_Instability_Deng_2017}.
It has been found that a correct treatment of charge conservation in the metal
bulk
and at its surface leads to striking differences: a potentially amplifying
channel related to non-specular scattering at the surface turns into 
damping when care is taken to avoid a charge sink at $z = 0$. The surface
current $J_z(0)$ that gives the major contribution in Deng's argument 
is actually absent
from the energy balance if the power exchanged between charges and field is
computed in an appropriate way. The final result for the imaginary part
of the surface plasmon frequency is
\begin{equation}
 \gamma = 
 - \frac{ 1 }{ 2\tau }
 - \left[ \frac{3}{8} + \frac{7}{96}(1-p) \right] 
 k v_{\mathrm{F}} 
 \,.
\label{eq:SCM_long_wave_imaginary_linear_dispersion}
\end{equation}  
In line with other publications using the semiclassical 
model\cite{Wagner_Oberfl_Wellen_e_Plasma,Zaremba_PhysRevB.9.1277}, 
we obtain an overall damped surface plasmon. Incidentally, 
non-specular scattering increases the damping proportional to the fraction $1-p$.
We shall see a similar result in Sec.\,\ref{subsec:disp_rel_HDM_GMF},
but obtained within a different approach that does not need to solve
the Boltzmann equation.
Eq.\,(\ref{eq:SCM_long_wave_imaginary_linear_dispersion}) is also consistent 
with a result of Zaremba using the SCM
\cite{Zaremba_PhysRevB.9.1277} as can be seen in his Table~III:
the damping increases progressively as the scattered electrons
have a more and more isotropic distribution.

\section{\label{sec:From_HDM_to_LDM}Hydrodynamic model}

In the following, we embed Deng's macroscopic model into an approach devised by Bedeaux and Vlieger (BV)\cite{Bedeaux_Vlieger_opt_prop_surf} which describes the charge density by a two-type model, as discussed in Sec.\,\ref{subsec:charge-conservation}. The model naturally accounts for the ``capacitive effect'' and illustrates why the continuity equation does not provide enough information to determine the dispersion relation. We then specialize to the hydrodynamic model (HDM), collect explicit formulas for field and current profiles and discuss two different ways to perform the local limit. This serves to clarify misconceptions of Deng about plasmons at specularly reflecting surfaces and about taking the local value of the surface current. This eventually leads to a natural interpretation why additional boundary conditions are absent in the local limit. Eventually, we will use a two-type charge model\cite{Horovitz_2012} to illustrate the role of a finite surface current.

\subsection{\label{sec:review_BV_embed_Deng_response}Bulk and surface charges}

In the BV approach, charge and current density are decomposed into
\begin{subequations}
\label{eq:Bedeaux_Vlieger_charge_current_decomposition}
\begin{eqnarray}
	{\rho}_{\rm tot}(\bm{x})   &=& \rho_{\rm s}\delta(z) + \rho(\bm{x})\Theta(z) \label{eq:Bedeaux_Vlieger_charge_decomposition}
\\
	{\bm{J}}_{\rm tot}(\bm{x}) &=& \bm{J}_{\rm s}\delta(z) + \bm{J}(\bm{x})\Theta(z)\,, 
\label{eq:Bedeaux_Vlieger_current_decomposition}
\end{eqnarray} 
\end{subequations}
where $\rho$ and $\bm{J}$ are restricted to the bulk metal ($z > 0$),
while the localized $\rho_{\rm s}$ as well as $\bm{J}_{\rm s}$ 
are called `excess quantities'\cite{Bedeaux_Vlieger_opt_prop_surf}. 
Analogous decompositions are applied for all fields. 
The excess quantities describe on a macroscopic scale the differences between the actual surface electrodynamics and the extrapolated bulk dynamics. If the excesses were absent, `a sharp transition from one bulk phase to the other' would be described, leading to Fresnel surfaces\cite{Bedeaux_Vlieger_opt_prop_surf}.

BV show that the normal component of the excess current $\bm{J}_{{\rm s},z}$ does not contribute to the matching conditions at the surface. So, without loss of generality, $\bm{J}_{\rm s}$ shall be directed along the surface. If we plug Eqs.\,(\ref{eq:Bedeaux_Vlieger_charge_current_decomposition}) into the continuity equation~(\ref{eq:continuity_equation}), separate localized and extended distributions and use the dependence of all fields proportional to $\exp[i(kx-\omega t)]$, we get
\begin{subequations}
\label{eq:BV_charge_conservation_surface_bulk}
\begin{eqnarray}
z>0: \qquad 0 &=& -{\rm i}\omega\rho(z) 
+ 
{\rm i} k {J}_x(z) + \partial_z {J}_z(z)
\label{eq:BV_charge_conservation_bulk}
\\
z=0: \qquad 0 &=& -{\rm i}\omega \rho_{\rm s}  
+ {\rm i} k {J}_{{\rm s},x} 
+ J_z(0^+)~.
\label{eq:BV_charge_conservation_surface}
\end{eqnarray}
\end{subequations}
which is the fixed-frequency representation of Eq.\,(\ref{eq:bulk-surface-charge-conservation}). We thus get a pair of continuity equations that are coupled by the bulk current $J_z(0^+)$ extrapolated to the surface. This current thus describes the charge exchange between bulk and surface.

As mentioned earlier, Deng does not split the charge density into surface and bulk parts. The parallel surface current $\bm{J}_{\rm s}$ is also absent in his model. This illustrates that already a certain ``additional boundary condition'' (ABC) has been applied: in his model, the electrons which accumulate at the surface are not allowed to move along it. 
Neglecting the parallel component of the surface excess current
(see Ref.\,\onlinecite{Flores_1972} for an estimation of its impact on
the surface plasmon dispersion),
Eq.\,(\ref{eq:BV_charge_conservation_surface}) yields
\begin{equation}
{\rm i}\omega \rho_{\rm s} = J_z(0^+)\,,
\label{eq:surface_charge_conservation_no_surface_divergence}
\end{equation}
as already used in Sec.\,\ref{ss:half-space}, Eq.\,(\ref{eq:continuity-charge-surface-current}).
A similar splitting of the charge into bulk and surface components can also
be spotted in Deng's papers. After Eq.(18) in Ref.\,\onlinecite{Possible_Instability_Deng_2017},
for example, the charge density computed from the distribution function is
identified as a bulk charge. The nonzero value of the surface current points
to a surface charge component, but the two components are not manifestly separated
in the cosine transform $\rho_q$ of the charge density. Within the approximation
$\rho_q \approx \rho_{\rm s}$, the bulk charge component would vanish.

The excess field formalism derived by BV illustrates that the continuity equation for the (total) charge density is not sufficient to determine the surface plasmon dispersion relation. In particular, Eq.\,(\ref{eq:surface_charge_conservation_no_surface_divergence}) must be supplemented by a model (often called an ABC) of how the accumulated charge reacts back on the surface current, e.g., by a repulsive force (``capacitive effect'') or a desorption process (``charge trap''). 
We provide a simple example in Sec.\,\ref{subsubsec:phenom_finite_surf_curr_HH_ABC}. 

\subsection{\label{sec:BV_formalism_application_to_HDM}Explicit sub-surface profiles}

In Secs.\,4.2 $\&$ A of Ref.\,\onlinecite{Deng_2019}, Deng discusses surface plasmons 
for the hydrodynamic electronic response. The latter is described by the (linearized) 
Euler equation of fluid dynamics which determines the current through
\begin{equation}
z > 0: \quad 
\bm{J}(z) = 
   \frac{\sigma }{1-{\rm i}\omega\tau}\bm{E}(z) 
 - \frac{v_0^2\tau }{1-{\rm i}\omega\tau}\nabla{\rho}(z)
 \,.
\label{eq:HDM_response_current}
\end{equation}
The first term is the Drude conductivity (DC value $\sigma$),
magnetic forces are neglected, since they
are of second order in deviations from equilibrium,
and the second one translates
the pressure arising from gradients in the charge density, using a 
linearized equation of state.
Its coefficient
is proportional to the compressibility of the electron fluid, 
and $v_0 = {\cal O}( v_{\rm F} )$
gives the speed of charge density waves in the bulk (longitudinal speed of sound).
The relaxation rate for the current density $1/\tau$ differs a priori
from the one for the distribution function, but we keep the same letter
for simplicity. 

The bulk density $\rho$ can be determined by virtue of Eqs.\,(\ref{eq:BV_charge_conservation_bulk},\ref{eq:HDM_response_current}).
Starting with an exponential \emph{Ansatz} and using Coulomb's law
$\mathop{\rm div} \bm{E} = 4\pi \rho$,
 one finds
\begin{equation}
\rho(z) = \rho(0^+) \, {\rm e}^{-\kappa z}
\,.
\label{eq:HDM_general_bulk_charge}
\end{equation}
The inverse (complex) length scale $\kappa$ is given by
\begin{equation}
\kappa^2 = k^2 
+ \frac{ \omega^2_{\rm p} - \omega\bar{\omega} }{ v_0^2 }~,
\label{eq:def-xi}
\end{equation}
where $\omega^2_{\rm p} = 4\pi \sigma / \tau$ is the squared plasma frequency.
Note the term $\omega\bar{\omega}$ which appears as $\bar\omega^2$ in Eq.\,(A7) 
of Ref.\,\onlinecite{Deng_2019} because of the wrong formulation of charge conservation.
Adopting the viewpoint that 
Eq.\,(\ref{eq:HDM_general_bulk_charge}) is the bulk charge and allowing
for a charge $\rho_{\rm s}$ localized at the surface, we find from 
Eqs.\,(\ref{eq:electrostatic_field_Greens_function}, \ref{eq:HDM_general_bulk_charge}) 
the electrostatic potential
\begin{subequations}
	\label{eq:HDM_complete_potential}
\begin{eqnarray}
z \ge 0: \qquad \phi(z) 
&=& \dfrac{2\pi}{k}\Big[ \rho_{\rm s} {\rm e}^{-kz} 
+ \rho(0^+)\Big( \dfrac{{\rm e}^{-\kappa z}}{k+\kappa} 
\nonumber \\
&& {} + \dfrac{{\rm e}^{-\kappa z} - {\rm e}^{-kz}}{k - \kappa}\Big)\Big]
\,,
\label{eq:HDM_metal_region_potential} 
\\
z \le 0: \qquad \phi(z) 
&=& \dfrac{2\pi}{k}\left( \rho_{\rm s} + \dfrac{\rho(0^+)}{k+\kappa}\right){\rm e}^{kz}
\label{eq:HDM_vacuum_region_potential}~.
\end{eqnarray}
\end{subequations}
From this and Eqs.\,(\ref{eq:HDM_response_current},
\ref{eq:HDM_general_bulk_charge}) the normal current density follows as
\begin{align}
z>0: \quad
& J_z(z) = 
\dfrac{\mathrm{i}}{\bar{\omega}} \bigg\{ \dfrac{\omega^2_{\rm p}}{2}\Big[\rho_{\rm s} {\rm e}^{-kz} + \dfrac{\rho(0^+)}{k}\Big(\dfrac{\kappa \, {\rm e}^{-\kappa z}}{k+\kappa} 
\nonumber \\
& \quad {} + \dfrac{\kappa \, {\rm e}^{-\kappa z} - k \, {\rm e}^{-kz}}{k - \kappa}\Big)\Big] + v_0^2\kappa \rho(0^+) {\rm e}^{-\kappa z}\bigg\}
\,.
\label{eq:HDM_normal_current}
\end{align}
These equations cannot be solved because one has to determine the ratio
$\rho_{\rm s} / \rho(0^+)$ between the two types of charges. 
To proceed, we adopt a model for the surface charge $\rho_{\rm s}$
or the surface current $J_z(0^+)$. The two are directly related because of the
charge conservation law~(\ref{eq:surface_charge_conservation_no_surface_divergence}).

\subsubsection{Usual hydrodynamic boundary condition} 
\label{sec:usual-HDM-BC}

In the hydrodynamic model, the condition 
\begin{equation}
J_z(0^+) = 0
\label{eq:HDM_zero_surface_current}
\end{equation}
can be interpreted as the impossibility of concentrating the electron fluid
into a true surface charge with zero extension at the surface; 
indeed, it implies $\rho_{\rm s} = 0$.
Because of
the charge gradient $\partial_z\rho(0^+) \ne 0$ in Eq.\,(\ref{eq:HDM_response_current}), 
this boundary condition 
does not lead to a vanishing surface field $E_z(0^+)$ 
and therefore allows for a surface plasmon mode\cite{Warren_Ferrell_PhysRev.117.1252},
as discussed by Ritchie and Gabor\cite{Plummer_et_al_impact_surface_plasmon}.

Deng claims in Appendix\,A of Ref.\,\onlinecite{Deng_2019} that this surface
plasmon mode is wrong, and we shall examine his arguments in parallel to
the relevant equations.
From Eqs.\,(\ref{eq:HDM_normal_current}, \ref{eq:HDM_zero_surface_current}),
we find [Ref.\,\onlinecite[Eq.\,(A9)]{Deng_2019}]
\begin{equation}
- \frac{\omega^2_{\rm s}}{\kappa + k} + {v_0^2}\kappa = 0
\label{eq:HDM_zero_current_xi_equation}
\end{equation}
with $\omega_{\rm s} = \omega_{\rm p}/\sqrt{2}$. 
We have assumed $\rho(0^+) \ne 0$.
Using the definition~(\ref{eq:def-xi}) of $\kappa$, one gets the dispersion 
relation
\begin{subequations}
\label{eq:HDM_zero_surface_current_disp_rel}
\begin{eqnarray}
\omega(k) &=& \Omega(k) - \dfrac{\mathrm{i}}{2\tau} 
\,,
\label{eq:HDM-complex-spdr}
\\
\Omega^2(k) &=& \omega^2_{\rm s} - \dfrac{1}{4\tau^2} 
+ k{v_0}\sqrt{ \omega^2_{\rm s} + \dfrac{k^2{v_0^2}}{4}} 
+ \dfrac{k^2 {v_0^2}}{2}
\,,
\nonumber\\
\Omega(k) & \approx & \omega_{\rm s} - \dfrac{1}{8\omega_{\rm s} \tau^2} 
+ \tfrac{ 1 }{ 2 } {v_0} k
\,.
\label{eq:HDM-spdr-small-k}
\end{eqnarray}
\end{subequations}
In the last expression, we have taken the small-$k$ limit to confirm
that $\omega_{\rm s}$ is the long-wavelength limit
of the surface plasmon dispersion, as it must\cite{Feibelman_1971,Flores_1972}. 
We recognize
that the small parameter of this expansion is $k{v_0}/\omega_{\rm s}$, 
which coincides with $k / k_s$ introduced by Deng\cite{Possible_Instability_Deng_2017,Deng_2019}.

Eqs.\,(\ref{eq:HDM-complex-spdr}, \ref{eq:HDM-spdr-small-k}) correspond to 
Eq.\,(A10) of Ref.\,\onlinecite{Deng_2019}, 
except that Deng obtains
a damping twice as large as here. Still, he claims that this surface
plasmon is ``plainly false''
because of its behaviour in the local limit, i.e., for $k{v_0}/\omega_{\rm s}
\ll 1$. His argument hinges on 
the limiting value of the
integrated (bulk) charge density $Q = \int\!{\rm d}z\, \rho(z)$.
This depends on two parameters. 
It is true, of course, that the spatial extent
$\sim 1/\kappa$ of $\rho(z)$ shrinks to zero,
as can be seen from Eq.\,(\ref{eq:def-xi}).
(We neglect, for the simplicity
of the argument, the imaginary part of $\omega$.) One thus gets
$Q = \rho(0^+)/\kappa \approx v_0 \rho(0^+) / \omega_{\rm s}$
in the local limit. For
a meaningful comparison, however, we have to express the
boundary value $\rho(0^+)$ by a quantity that is well-defined 
in this limit. One candidate is indeed the potential $\phi(0)$: 
from Eq.\,(\ref{eq:HDM_complete_potential}) with $\rho_{\rm s}=0$,
we find
\begin{equation}
\rho(0^+) = \frac{k\phi(0)}{2\pi}(k+\kappa)~.
\label{eq:amplitudes_charge_density_potential}
\end{equation}
Pulling these two expressions
together, we have 
\begin{equation}
Q = \int\!{\rm d}z\, \rho(z) \approx \frac{ k \phi(0) }{ 2\pi }
\quad \text{for } \kappa \to \infty
\,.
\label{eq:macroscopic-HDM-charge}
\end{equation}
This coincides with 
the jump in the normal electric field, as the potential approaches in 
the local limit the form $\phi(z) \approx \phi(0)\,{\rm e}^{ - k |z| }$
from Eqs.\,(\ref{eq:HDM_complete_potential}). 
Deng's argument that the integrated charge density vanishes in the local 
limit\cite{Deng_2019} is thus fallacious.

\subsubsection{The local limit}

We have seen that in the local limit, the charge density of the hydrodynamic
model shrinks to a surface charge. How does the current density behave to avoid
a conflict between the boundary condition~(\ref{eq:HDM_zero_surface_current})
and the charge conservation law Eq.\,(\ref{eq:surface_charge_conservation_no_surface_divergence})? The answer
requires elements from boundary layer (or multiple scale) 
techniques\cite{Nayfeh_Intro_Pert_Techn}, since
the sub-surface region where the current drops to zero is shrinking to an
infinitely thin layer as $v_0 \to 0$.
In this paragraph, we stick to the case $\rho_{\rm s} = 0$ and work out
how the \emph{bulk} charge density $\rho( z )$ \emph{apparently} becomes
localized when the local limit is taken.

\begin{figure}
	\includegraphics[width=0.77\columnwidth]{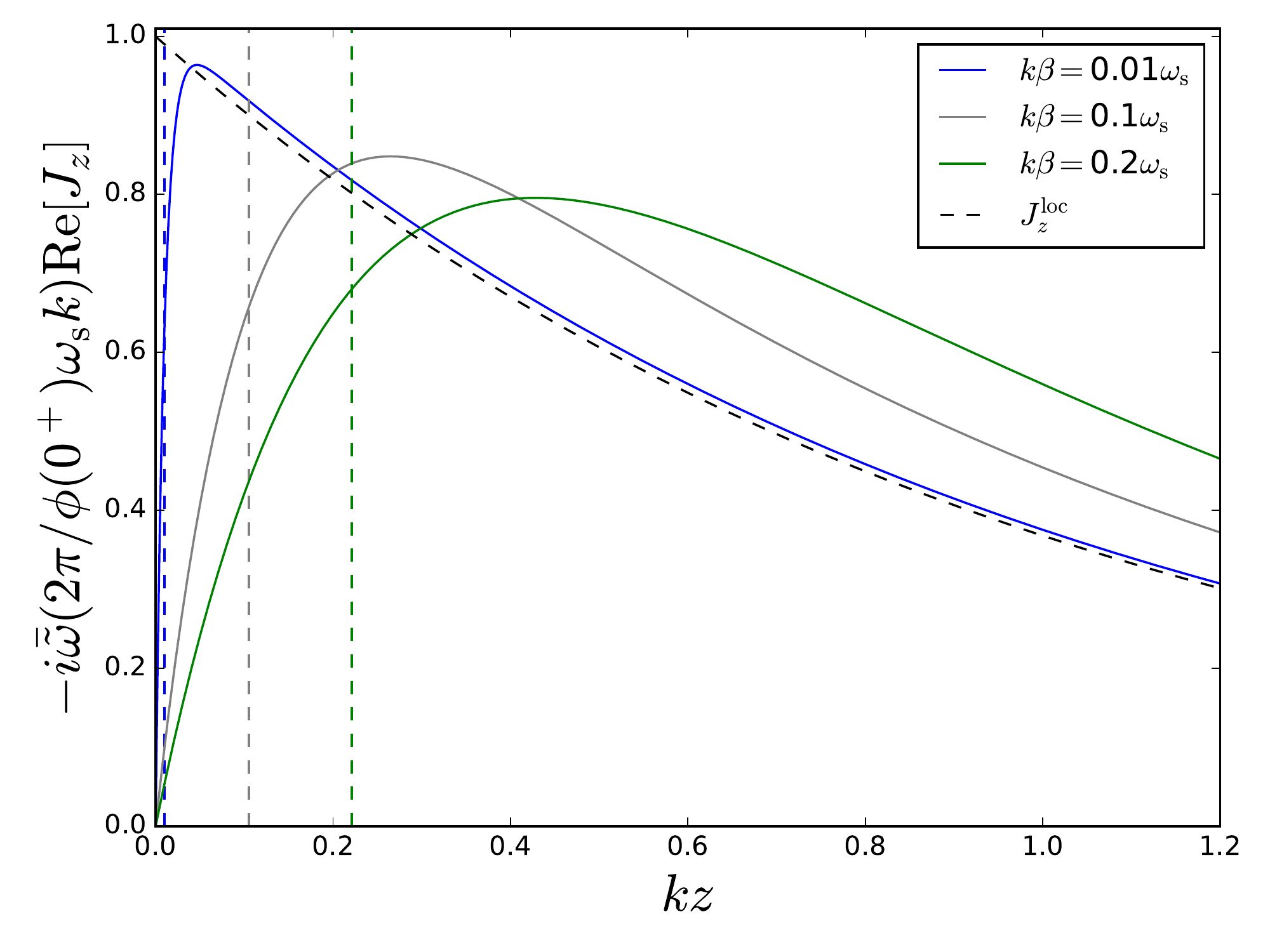}
	\caption{\label{sec:HDM_zero_surface_current_analytic_solution_outer_expansion} 
Illustration of the normal current given analytically 
by Eqs.\,(\ref{eq:HDM_normal_current}) $\&$ (\ref{eq:HDM_zero_current_xi_equation}) 
within the metal volume for decreasing values of $k{v_0}/\omega_{\rm s}$ 
in comparison with the usual local solution (outer expansion) given by 
Eq.\,(\ref{eq:usual_local_limit_normal_current}). }
\end{figure}

Let us note first that when the limit $v_0 \to 0$ is taken in
Eq.\,(\ref{eq:HDM_response_current}), the order of the differential equation
is reduced, which changes qualitatively the number of boundary conditions.
Since the Maxwell boundary conditions for the fields hold independently
of the local limit, it is the ABC~(\ref{eq:HDM_zero_surface_current}) that
has to be discarded.
The behavior of the current density is illustrated in 
Fig.\,\ref{sec:HDM_zero_surface_current_analytic_solution_outer_expansion}
for different values of the parameter $k v_0 / \omega_{\rm s}$. Note
how the current profile degenerates into a jump at the surface as
$k v_0 / \omega_{\rm s} \to 0$. In boundary layer 
theory, one introduces
an ``inner expansion'' for $0 \le z \alt v_0 / \omega_{\rm s}$ (marked by the
vertical dashed line) and an ``outer'' one for $z \sim 1/k$. 
The two length scales get widely different in the local limit. 

The outer expansion describes the current on the macroscopic scale.
We keep $z > 0$ fixed in Eq.\,(\ref{eq:HDM_normal_current})
and take the limit $v_0 \to 0$ or equivalently $\kappa \to \infty$,
giving
\begin{equation}
z > 0: \quad
\lim\limits_{\kappa \to \infty} J_z(z) 
= \frac{\mathrm{i}\omega_{\rm p}^2}{4\pi \bar{\omega} }
k\phi(0) {\rm e}^{-kz}
~. 
\label{eq:usual_local_limit_normal_current}
\end{equation}
In this calculation, we discard terms ${\rm e}^{ - \kappa z} \to 0$ and
keep ${\rm e}^{ - k z}$. This is the well-known local response according
to Ohm's law with an AC conductivity $\sigma(\omega) 
= 
\omega_{\rm p}^2 \tau / (4\pi ( 1 - {\rm i} \omega \tau))$
[see Eq.\,(\ref{eq:HDM_response_current})].
This is shown as the dashed curve in 
Fig.\,\ref{sec:HDM_zero_surface_current_analytic_solution_outer_expansion}.
Since the outer expansion discards the boundary layer, it is not surprising
that taking formally the limit $z \to 0$ in 
Eq.\,(\ref{eq:usual_local_limit_normal_current}) gives
a nonzero surface current. Its value conforms with 
charge conservation~(\ref{eq:surface_charge_conservation_no_surface_divergence}),
considering the integrated
charge density~(\ref{eq:macroscopic-HDM-charge}) \ldots provided
the frequency is fixed to $\omega \bar{\omega} = \omega_{\rm s}^2$!
In this way, the local calculation recovers the long-wavelength
dispersion relation [including the losses spelled out 
in Eq.\,(\ref{eq:HDM_zero_surface_current_disp_rel})]. 
A glance at Fig.\,\ref{sec:HDM_zero_surface_current_analytic_solution_outer_expansion}
illustrates how the zero-surface-current ABC cannot be satisfied in the
local limit.

To illustrate the characteristic behavior of the current within the boundary layer, 
we display the inner expansion. We work on the short length scale and take
$z \sim 1/\kappa$, where the hydrodynamic pressure (i.e, the density gradient)
is significant, while the condition $k z \ll 1$ expresses the separation of
length scales in the local limit. 
Using Eqs.\,(\ref{eq:HDM_normal_current}, \ref{eq:amplitudes_charge_density_potential}), 
we find
\begin{equation}
z \sim 1/\kappa: \quad
\lim\limits_{1/k \to \infty} 
J_z(z) = k \phi(0) \sigma( \omega ) 
\left( 1 - {\rm e}^{-\kappa z}\right)
~.
\end{equation}     
We recognize that this current suits the condition of a 
vanishing surface current and saturates to its local limit
for $1/ \kappa \ll z \ll 1/k$.

\subsection{\label{sec:Deng-critique-HDM}Critique of the 
hydrodynamic surface plasmon}

In the two previous subsections, we have tried to argue how the
surface plasmon of the {usual} hydrodynamic calculation connects smoothly with
the local limit. We learned that the boundary
condition~(\ref{eq:HDM_zero_surface_current}) for the surface current
does not conflict with the emergence of a surface mode, quite distinct
from the claims in Refs.\,\onlinecite{Universal_Self_Amplification_Deng_et_al_2017,
Possible_Instability_Deng_2017, Deng_2017b_metal_film, Deng_2019}.

\subsubsection{Dielectric function}

The condition of a vanishing surface current is often interpreted 
as describing specular scattering at the surface.\footnote{This is not quite correct. Recall that the boundary 
condition $J_z(0) = 0$ has also been used in
the Boltzmann solution of Sec.\,\ref{sec:SCM} using a
Fuchs parameter $0 < p < 1$.} 
The specular
scattering case $p = 1$ is special in the sense that a dispersion relation 
can be found with the help of a symmetry argument, using any dielectric
function (conductivity) in the bulk 
metal \cite{Wagner_Oberfl_Wellen_e_Plasma, Zaremba_PhysRevB.9.1277,
GMF_intro_theory_solid_surfaces}. This relation reads
\begin{equation}
1 = \frac{ k }{ \pi } \int\!\frac{ {\rm d}q }{ K^2 \varepsilon_L( K, \omega ) }
\,,
\label{eq:specular-dispersion-relation}
\end{equation}
where $\varepsilon_L( K, \omega )$ is the bulk dielectric function that
only depends on the modulus $K$ of the wave vector
$\bm{K} = (k, 0, q)$. The integral~(\ref{eq:specular-dispersion-relation}) 
can be worked out analytically as a contour integral 
by taking the
(longitudinal) dielectric function of the hydrodynamical model
\begin{equation}
\mbox{HDM:}\qquad
\varepsilon_L( K, \omega ) 
= 1 - \frac{ \omega_{\rm p}^2 }{ \omega \bar{\omega} - v_0^2 K^2 }
\,.
\label{eq:HDM-dielectric-function}
\end{equation}
The resulting sum over residues then yields the dispersion 
relation~(\ref{eq:HDM_zero_surface_current_disp_rel})
found before by an elementary calculation.

We note that the dielectric function~(\ref{eq:HDM-dielectric-function})
corresponds to Eq.\,(\ref{eq:HDM_response_current}). This can be checked
via the conductivity $\sigma_L( K, \omega )$. (The subscript $L$
is for ``longitudinal'', as we are dealing with an electric field
parallel to $\bm{K}$ in Fourier space.)
Taking the spatial Fourier 
transform of~(\ref{eq:HDM_response_current}) in a bulk medium 
and noting that $\bm{K} \cdot \bm{J} = \omega \rho$ 
(charge conservation), one gets
$\bm{J} = \sigma_L( K, \omega ) \bm{E}$ with
\begin{equation}
\sigma_L( K, \omega ) 
= \frac{ {\rm i} \omega \sigma / \tau }{ \omega \bar{\omega} - v_0^2 K^2 }
\,.
\label{eq:HDM-conductivity}
\end{equation}
The standard relation
$\varepsilon_L( K, \omega ) = 1 + 4 \pi {\rm i} \sigma_L( K, \omega ) / \omega$
then yields the hydrodynamic model~(\ref{eq:HDM-dielectric-function}).
The pole of Eqs.\,(\ref{eq:HDM-dielectric-function}, 
\ref{eq:HDM-conductivity}) at $\omega \approx v_0 K$ 
is characteristic for the intrinsic sound waves in the electron gas.
The zeros of the dielectric function~(\ref{eq:HDM-dielectric-function})
corresponds to bulk plasma waves that approach $\omega \to \omega_{\rm p}$
for $k \to 0$ and disperse
$\sim v_0 K$ at large $K \gg \omega_{\rm p} / v_0$. Replacing $1$ by
a background dielectric constant would be a way to incorporate the response
of bound charges (e.g., d-electrons in Gold).

Deng claims that
there is a ``non-equivalent approach'' to the hydrodynamical model and
uses for the dielectric function [Sec.\,4.2 and Footnote\,1 in 
Ref.\,\onlinecite{Deng_2019}]
\begin{equation}
\text{Deng:}\quad
\varepsilon_L( K, \omega ) 
= 1 - \frac{ \Omega_0^2 + v_0^2 K^2 }{ \bar{\omega}^2 }
\,,
\label{eq:Deng-dielectric-fcn-HDM}
\end{equation}
where $\Omega_0$ must be identified with the plasma frequency $\omega_{\rm p}$
from the small-$K$ and high-frequency asymptotics. The condition 
$\varepsilon_L = 0$ gives a similar dispersion relation for bulk plasma
waves as above (with a different damping because of the square $\bar{\omega}^2$
in the denominator), but sound-wave poles are absent. 
Note that the exact factors $\omega$ 
in Eqs.\,(\ref{eq:HDM-dielectric-function}) and~(\ref{eq:HDM-conductivity})
arise from charge conservation and the relation $\bm{J} = - {\rm i} \omega \bm{P}$
between the current density and the polarization field.

\subsubsection{Breaking of translation symmetry}
\label{sss:discussion-boundary-terms}

The fundamental equation for the surface plasmon dispersion
in Refs.\,\onlinecite{Universal_Self_Amplification_Deng_et_al_2017,
Possible_Instability_Deng_2017, Deng_2017b_metal_film, Deng_2019}
is Eq.\,(\ref{eq:Deng_def_surface_diel_fctn}) above,
where the integral operators
${\cal H}$ and ${\cal G}$ introduced in 
Eqs.\,(\ref{eq:Deng_surface_model_continuity_with_integral_operator},
\ref{eq:def-G-kernel}) appear. 
What attracts attention is that for the bulk ${\cal H}$ operator,
in most cases only the translation invariant part is taken, i.e.
the kernel is approximated ${\cal H}( q, q' ) \sim \delta( q - q' )$
[see Eq.\,(\ref{eq:Deng_macroscopic_response_volume_operator_eigenvalue_Fourier})
above].
Within the hydrodynamic formula~(\ref{eq:HDM_response_current}) 
for the current density, we can identify the correction to this
approximation. 
After a partial integration, one finds from the 
definition for the kernel ${\cal H}$:
\begin{align}
& \int dq'\, \mathcal{H}(q,q') \rho(q') =
\int\limits_{0}^{\infty}\!{\rm d}z \cos(q z)
(- {\rm i} \bar{\omega}) \nabla \cdot \bm{J}
\nonumber\\
&= 
v_0^2 \partial_z\rho(0^+)
+ 
\int_{0}^{\infty} dz\,\cos(q z)
\left( \omega^2_{\rm p} + v_0^2K^2 \right) \rho(z)
\label{eq:Kernel_H_HDM_step1}
\end{align}
where 
we have set
$K^2 = k^2 + q^2$. The integral on the rhs gives the cosine transform
$\rho_q$ and represents the bulk (translation invariant) kernel
${\cal H}_{\rm b}( k, q )
= \left(\omega^2_{\rm p} + v_0^2K^2\right) \delta(q-q')$.
{We shall argue in the following that 
the other term in ${\cal H}$ is nonzero, so that
Deng's statement that boundary terms are generally negligible 
compared to $\mathcal{H}_{\rm b}$, should be treated with caution.}

{It is the derivative
$\partial_z\rho(0^+)$ that breaks translation invariance.} It is
set to zero by Deng because of the general form of the cosine 
transform~(\ref{eq:definition_cosine_transform})
[see, e.g., Ref.\,\onlinecite{Deng_2019} before Eq.\,(A13)].
This is a subtle point, in particular when the local limit is considered.
Indeed, the exponential charge density~(\ref{eq:HDM_general_bulk_charge})
does have a nonzero derivative $\partial_z\rho(0^+) = - \kappa \rho(0^+)$.
In the local limit $\kappa \to \infty$,
this charge density provides an example of a ``skew'' representation
of the $\delta$-function that is entirely localized in the region
$z \ge 0$.
Its cosine transform is well-defined
for $\kappa < \infty$ and provides the integral representation
\begin{equation}
\rho(z) = \Theta(z) \frac{ 2Q \kappa^2 }{\pi}
\int_0^{\infty}\!{\rm d}q \, \frac{ \cos(qz) }{ \kappa^2 + q^2 }
\label{eq:cosine-expansion-rho-z}
\end{equation}
Here, $Q$ is the integrated charge (per area) defined in
Eq.\,(\ref{eq:macroscopic-HDM-charge}).
The evaluation of $\partial_z \rho$ requires
an UV regularization of the integral. 
This can be performed by evaluating the integral as a contour integral
along the entire real line in the complex $q$-plane, writing the integrand as 
$q \, {\rm e}^{ iqz } /( \kappa^2 + q^2 )$, closing the contour with a 
circle at infinity in the upper half-plane  
and picking the
residue at $q = i \kappa$. If the limit $\kappa \to \infty$ is 
taken first 
in Eq.\,(\ref{eq:cosine-expansion-rho-z}), one gets a ``symmetric'' 
$\delta$-function and the prefactor $\Theta(z)$ halves its weight
to be consistent with the real-space representation $\rho(z)$.
{Keeping $\kappa$ finite, on the other hand, one gets a function
$\sim {\rm e}^{ - \kappa |z| }$ whose derivative at $z = 0$ is not zero,
but shows a jump. This is in stark contrast to the naive analysis of
the integrand around $z = 0$. (For more details on these integrations,
see Appendix~\ref{as:cos-and-exp-integrands}.)}

We now evaluate the kernel ${\cal G}(k, q)$ that links the surface 
current $J_z(0^+)$ to the charge density $\rho_q$ [Eq.\,(\ref{eq:def-G-kernel})].
From Eq.\,(\ref{eq:HDM_response_current}), we get
\begin{equation}
{\rm i}\bar{\omega} J_z(0^+) = \frac{ \omega_{\rm p}^2 }{ 4\pi } 
\partial_z \phi(0) + v_0^2 \partial_z \rho(0^+)
\label{eq:surface-current-HDM}
\end{equation}
For the potential~$\phi(0)$ in the first term, there is no need to
make the limit $z \downarrow 0$ explicit if we may assume that it is
continuous across the boundary. (Only a ``double layer'' or perpendicular
surface polarization would change this picture.) This term 
corresponds to the Drude model and would be taken by Deng as the
translation-invariant part $G_{\rm b}$. 
One may wonder whether the second term with the derivative 
$\partial_z \rho(0^+)$ breaks translation invariance: after all, 
the
derivative $\partial_z\rho$ appears in the same form anywhere in the bulk.
This may have
lead Deng to the statement
[Ref.\,\onlinecite{Deng_2019}, Sec.\,4.2 preceding Eq.\,(21)] 
that symmetry breaking is absent in the hydrodynamic model.

We would like to put forward the viewpoint that the symmetry breaking 
arises in hydrodynamics from the boundary condition $J_z(0^+) = 0$ itself.
From this viewpoint,
the hydrodynamic surface plasmon appears related to a charge distribution
in the \emph{kernel} of the non-trivial integral operator 
\begin{eqnarray}
\mbox{HDM:}\qquad
{\cal G}(k, q) & = & -\frac{ \omega_{\rm p}^2 }{ \pi } 
\lim_{z \downarrow 0} \left( 2 q \sin(q z) - k \right)
\nonumber\\
&& {}
- \frac{ 2 v_0^2 }{ \pi } \lim_{z \downarrow 0} q K^2 \sin(q z)
\label{eq:}
\end{eqnarray}
where the definition~(\ref{eq:def-G-kernel}) of the integral operator
${\cal G}$ was applied.
The first term corresponds to the Ohmic current response to the electric
field, the second term to the charge density gradient.
Note that the limit $z \downarrow 0$ must be taken as the last step [after the $q$-integration of Eq.\,(\ref{eq:def-G-kernel})],
since one may have to deal with singular charge distributions where
values $q \to \infty$ are significant. The local limit $\rho( z )
\to \rho_{\rm s} \delta( z )$ provides an 
example. Deng ignores the contribution $2 q \sin(q z)$ to the first
term and finds an electric field with the opposite sign. (It would
apply at $z \to 0^-$ rather than $z \to 0^+$ for a localized charge,
see Eq.\,(\ref{eq:electric_field_displaced_charge}).)
One may speculate whether such a sign change may be responsible for
the surface plasmon amplification, when charges seem to ``flow uphill''.

In view of this discussion, we may comment on the status of 
Eq.\,(\ref{eq:Deng_def_surface_diel_fctn}). It is fundamental to Deng's
analysis, see e.g., the evaluation in Ref.\,\onlinecite{Deng_2019}, where a
splitting $G = G_b + G_s$ is done and the imaginary part of $G_s$ 
is responsible for amplification. The
denominator~$\Omega^2(k,q) - \bar{\omega}^2$ in 
Eq.\,(\ref{eq:Deng_def_surface_diel_fctn}) arises from the bulk approximation
to the kernel ${\cal H}$ as mentioned above, and misses the surface correction
discussed above. 
If in the derivation one simplifies by $J_z(0) \ne 0$, 
then by charge conservation, a genuine
surface charge $\rho_{\rm s}$ must be present, which is distinct, 
however, from the volume charge of which $\rho_q$ is the cosine transform. 
We discuss such a ``composite model'' in the following section.

\subsection{\label{subsubsec:phenom_finite_surf_curr_HH_ABC}Composite charge model}

To elaborate on the nature of the surface charge and current, we want to introduce 
a phenomenological boundary condition which has been derived by Horovitz and 
Henkel \cite{Horovitz_2012} for this mesoscopic model. Solving the 
Boltzmann equation for the volume and (infinitely thin) surface region separately, 
but allowing for a collision term that mixes surface and bulk electrons, they obtain 
\begin{equation}
J_z(0^+) = \frac{\rho_{\rm s}}{\tau_{\rm d}} - \alpha{v_0}\rho(0^+)\,,
\label{eq:HH_ABC}
\end{equation}
where $1/\tau_{\rm d}$ and $\alpha$ describe 
the desorption rate of surface electrons back into the bulk
and a probability of trapping a bulk electron in the surface, 
respectively.\cite{Horovitz_2012} The desorption process prevents the unphysical 
accumulation of charges at the surface which we have discussed in 
Sec.\,\ref{subsec:charge-conservation}. 

Eq.\,(\ref{eq:HH_ABC}) provides 
the essential information on how the mixing of surface and bulk electrons takes place. 
This could not be accounted for by the continuity equation of the whole system itself, 
as we have argued after Eq.\,(\ref{eq:surface_charge_conservation_no_surface_divergence}).
If we had assumed a vanishing surface current, then Eq.\,(\ref{eq:HH_ABC})
would describe the balance of trapping and desorption. But then
also $\rho_{\rm s}$ would vanish by
Eq.\,(\ref{eq:surface_charge_conservation_no_surface_divergence}).
Hence, to allow at least for a nonzero bulk charge, we would be forced to
take $\alpha = 0$ and mixing between the two charges would be completely absent. 

In the following, we assume $J_z(0^+) \ne 0$ which may be interpreted
as a measure of non-specular scattering.
Combined with charge conservation, we then get
\begin{equation}
\rho_{\rm s} = \frac{ \alpha v_0 \tau_{\rm d} }{ 1 - {\rm i} \omega \tau_{\rm d}} \rho(0^+)
\label{eq:ratio-of-charges}
\end{equation}
Using Eqs.\,(\ref{eq:HH_ABC}, \ref{eq:ratio-of-charges}) in Eqs.\,(\ref{eq:surface_charge_conservation_no_surface_divergence}, \ref{eq:HDM_normal_current}) we find the implicit dispersion relation
\begin{eqnarray}
0 &=& 
      - \dfrac{ \omega^2_{\rm s} }{k + \kappa}
      + v_0^2\kappa 
\nonumber \\
&& {} + \frac{ \alpha {v_0} \tau_{\rm d}}{ 1 - {\rm i} \omega \tau_{\rm d} }
\left[ 
\omega^2_{\rm s} - \omega\left(\omega + \dfrac{\mathrm{i}}{\tau}\right)
\right]
\,.
\label{eq:HH_ABC_disp_rel_implicit_dimensional}
\end{eqnarray}
The solution in the local limit $v_0 \to 0$ is given by the well-known 
surface plasmon frequency \cite{Feibelman_1971, Flores_1972}, i.e. 
\begin{equation}
{v_0} \to 0 : \quad \omega \to 
\sqrt{\omega^2_{\rm s} - \frac{1}{4\tau^2}} - \frac{\mathrm{i}}{2\tau}
\,.
\end{equation}
In particular, trapping and desorption at the surface are irrelevant on the local 
scale -- in contrast to Eq.\,(51) and Fig.~A1(a) of Ref.\,\onlinecite{Deng_2019} 
where the plasmon resonance depends on surface scattering.

\begin{figure}
	\includegraphics[width=0.83\columnwidth]{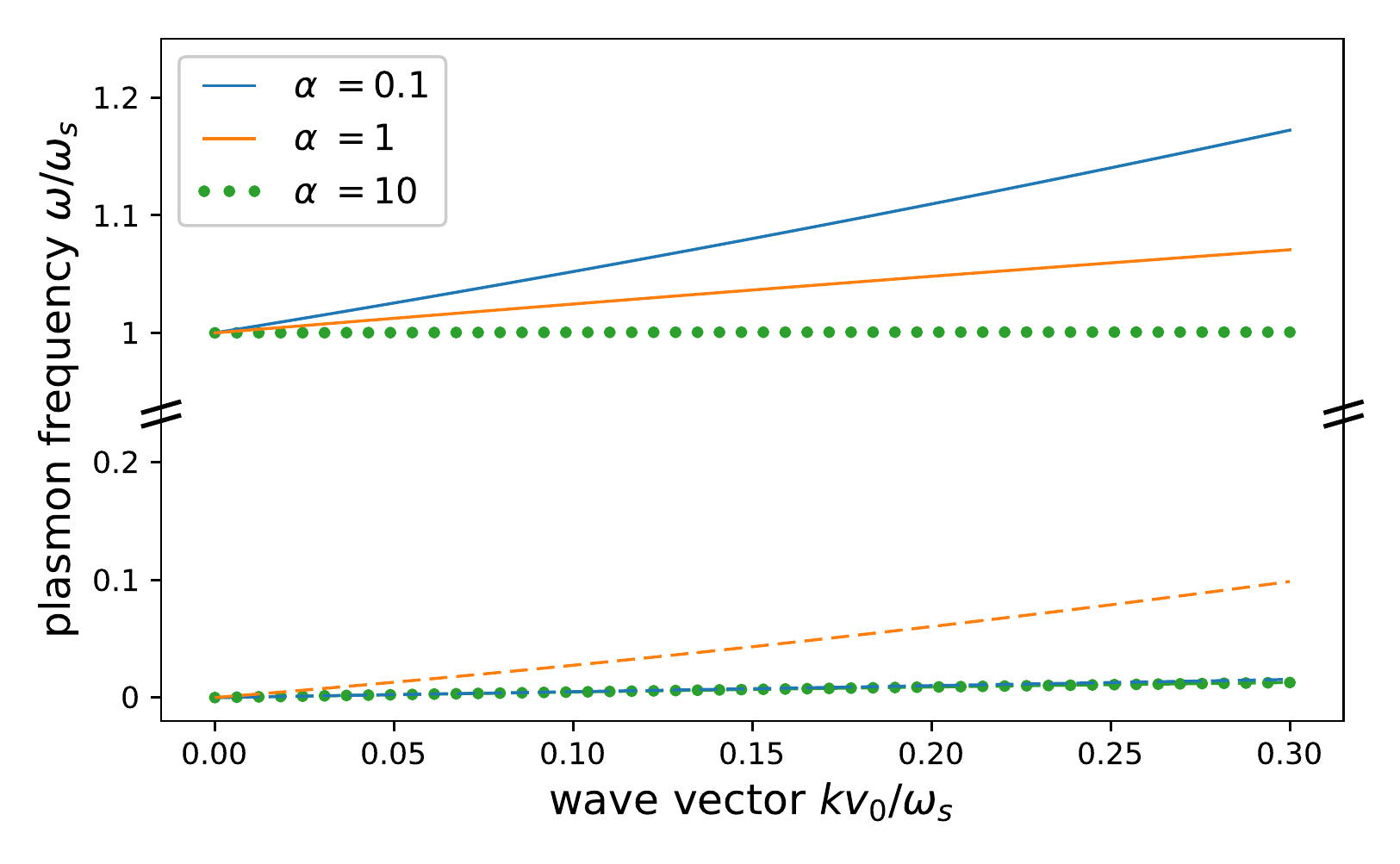}
	\caption{Surface plasmon dispersion relation $\omega( k )$ in the hydrodynamic model
for a composite charge density (partly localized at the surface, partly
in the sub-surface region). We plot the results of 
Eqs.\,(\ref{eq:HH_ABC_disp_rel_asympt_ansatz}, 
\ref{eq:HH_ABC_disp_rel_higher_dispersions}),
normalized to the local limit $\omega_{\rm s}$
for a desorption rate $1/\tau_{\rm d} = 0.01\, \omega_{\rm s}$) 
and different values for the trapping parameter $\alpha$.
Solid lines denote the real part and dashed lines the negative 
imaginary part.}
\label{fig:HH_ABC_disp_rel}\end{figure}

We are interested in solving Eq.\,(\ref{eq:HH_ABC_disp_rel_implicit_dimensional}) 
for small $k$ and expand in 
powers of $k{v_0}/\omega_{\rm s}$. For simplicity, bulk collisions are neglected
($\tau \to \infty$). Introducing the dimensionless quantities
\begin{equation}
\tilde{\omega} = \frac{\omega}{\omega_{\rm s}}
\,, \quad
\tilde{v}_0 = \frac{k{v_0}}{\omega_{\rm s}}
\,, \quad
\tilde{\kappa} = \frac{ \kappa {v_0} }{\omega_{\rm s}}
\,, \quad 
\text{and} \quad
\tilde{\tau}_{\rm d} = \omega_{\rm s}\tau_{\rm d}
\,,
\end{equation} 
we transform Eq.\,(\ref{eq:HH_ABC_disp_rel_implicit_dimensional}) into 
dimensionless form
\begin{equation}
0 =
- \frac{1}{ \tilde{\kappa} + \tilde{v}_0 } + \tilde{\kappa} 
+ 
\alpha \tilde{\tau}_{\rm d}
\frac{ 1 - \tilde{\omega}^2 }{ 1 - \mathrm{i} \tilde{\omega} \tilde{\tau}_{\rm d} }
\,.
\label{eq:HH_ABC_disp_rel_implicit_dimensionless}
\end{equation}
An expansion in powers of $\tilde{v}_0$ gives the dispersion relation as
\begin{equation}
\omega( k ) 
= \omega_{\rm s} + a \, k v_0  + b \, \frac{ (k v_0)^2 }{ \omega_{\rm s} }
\,.
\label{eq:HH_ABC_disp_rel_asympt_ansatz}
\end{equation}
Expanding $\tilde{\kappa}$ from Eq.\,(\ref{eq:def-xi}) 
to second order in $\tilde{v}_0$,
and equating like powers of $\tilde{v}_0$, we find to the second order
\begin{subequations}
\label{eq:HH_ABC_disp_rel_higher_dispersions}
\begin{eqnarray}
a &=&
\dfrac{1 
	- \mathrm{i}\alpha
	+ \alpha / \tilde{\tau}_{\rm d}
	+ 1 / \tilde{\tau}^2_{\rm d}
}{
2 + 2 \left(\alpha + 1 / \tilde{\tau}_{\rm d}\right)^2
} 
\label{eq:HH_ABC_disp_rel_linear_dispersion} 
\\
b &=& \dfrac{
a \left( 1 + \mathrm{i}/{\tilde{\tau}_{\rm d}} \right)
\left( 2 - 3 a \right) 
+ a \left( 1 - \mathrm{i} \alpha - 2 a \right) 
}{
1 + 2\mathrm{i}\alpha + \mathrm{i}/ \tilde{\tau}_{\rm d}
}
\,.
\label{eq:HH_ABC_disp_rel_quadratic_dispersion}
\end{eqnarray}
\end{subequations}
These results are plotted in Fig.\,\ref{fig:HH_ABC_disp_rel}. 

From Refs.\,\onlinecite{Harris_1971a, Flores_1972, Feibelman_1989}, 
we know that the linear dispersion of the real part of $\omega(k)$ 
(Fig.\,\ref{fig:HH_ABC_disp_rel}, upper set of curves)
is proportional to the centroid of the charge density.\footnote{The centroid is defined as $\int\!dz\, z \rho( z )$;
the origin being chosen such that the difference
between the equilibrium charge $\rho_0(z)$ and a step function 
placed at $z = 0$ integrates to zero.\cite{Harris_1971a}}
It vanishes for a pure surface charge and increases with the ratio of volume
to surface charge. From Eq.\,(\ref{eq:ratio-of-charges}), this case
corresponds to small $\alpha$, consistent with the Figure.
The damping (lower curves)
is always regular (no amplification), and the maximum of its coefficient linear
in $k$ is achieved at $\alpha=1$ 
when we consider the limit $\omega_{\rm s} \tau_{\rm d} \to \infty$
in Eq.\,(\ref{eq:HH_ABC_disp_rel_linear_dispersion}).
This damping may be attributed to the charge trapping in the surface layer.

These results are consistent with Ref.\,\onlinecite{Horovitz_2012} and
obtained within a simpler calculation. One difference is that the model
considered here
cannot give a negative linear dispersion. This is related to the approximation 
that $\rho_{\rm s}$ represents a true surface charge and that one takes 
the position $z = 0$ as a reference for the charge centroid.

We conclude that this hydrodynamic model of a composite charge distribution
incorporates all elements of Deng's approach: the surface breaks translation
invariance, the nonzero surface current describes non-specular surface scattering
 (similar to Fuchs' $p$-parameter), 
there is a surface plasmon mode with a well-defined local
limit, the charge trapping at the surface reflects the ``capacitive effects'' 
highlighted in Ref.\,\onlinecite{Deng_2019} -- and still this 
model predicts a surface plasmon which is damped rather than unstable. 
The differences
with respect to the dispersion found in the semiclassical approach (Boltzmann 
equation, Sec.\,\ref{sec:SCM}) can be attributed to the different dynamics
involved in surface scattering: the Zaremba prescription in the semiclassical
model gives an ``instantaneous desorption'' of non-specularly scattered charges,
while here, the parameter $\tau_{\rm d}$ plays the role of a mean dwell time.

\section{\label{sec:pseudo_model}{Extended-medium} approach}

In this section, we review another macroscopic approach to the electromagnetic
response of surfaces that incorporates surface roughness and has been introduced 
by Garc\'{i}a-Moliner and Flores (GF)\cite{GMF_intro_theory_solid_surfaces}. 
We shall dub it the ``pseudo-model'' in the following. 
Its application in Ref.\,\onlinecite{GMF_cond_Surfaces_non_specular} deals with 
almost the same problem as Deng \cite{Deng_2019}. In particular, it is independent 
of the particular choice of the electron dynamics (bulk dielectric function). 
The construction is based on the fields rather than the charge or current density. 
Another difference is the restriction to a vanishing surface current (no physical
surface charge).

We review the approach, discuss the nature of the corresponding surface plasmons 
and then propose connections 
to the specular reflection model and the dielectric approximation.

\subsection{\label{subsec:Pseudomodel} Physical half-spaces and fictitious stimuli}

To describe the response of a metallic half-space, GF distinguish
between two classes of paths that a charge can take in the 
medium to reach a given point \cite{GMF_intro_theory_solid_surfaces}. 
As discussed in the Boltzmann theory (Sec.\,\ref{sec:SCM}), 
one class describes ``direct propagation''  and is only determined by 
the metal's bulk properties, in particular it is translationally invariant. 
Apart from that, there are also paths that touch the surface and are
scattered there. To model this, both the metal and the vacuum half-spaces 
are augmented by the other half-space. Of course, the new half-space is fictitious
in nature. The extended media will be called `pseudo-media'. 
Now, to simulate surface effects, the actual perturbation at some point 
$\bm{x}'=(x,y,z>0)$ will be mimicked by its mirror image at 
$\bm{x}'_-=(x,y,-z)$. From there, a charge is assumed to propagate through the 
pseudo-medium towards $\bm{x}$ without further perturbation
by the surface. Say we are concerned with the current $\bm{J}$ within the metal. 
Then, the constitutive relation is given by
\begin{equation}
\bm{J}(z) = \int_{z'>0} dz'~ \left[ \sigma(\bm{x}-\bm{x}') \bm{E}(\bm{x}') + \sigma(\bm{x}-\bm{x}'_-) \bm{E}(\bm{x}'_-)\right]~.
\label{eq:pseudomodel_constitutive_eq_without_p_model}
\end{equation}
Note that this formulation does not yet fix the value of the surface current. 
This requires first the determination of the field $\bm{E}$ in the entire pseudo-medium.
If this field depends linearly on the physical field in the metal half-space,
the second term in Eq.\,(\ref{eq:pseudomodel_constitutive_eq_without_p_model})
can be identified with the symmetry-breaking conductivity
$\sigma_{\rm s}( z, z' )$ introduced by Deng 
[Eq.\,(\ref{eq:Deng_decomposition_conductivity})].

The extended-medium model is completed in three steps. First of all, GF devise a set of ``fictitious stimuli''. These are currents and charges which are placed outside the respective real medium. They generate the field in the whole pseudo-metal. 
(Note that the approach is applied with electric and magnetic fields and allowing
for externally incident fields.)
If the field $\bm{E}(\bm{x}'_-)$ is constructed by a mirror symmetry, its normal
component shows an (unphysical) jump at the surface which has to be compensated by
introducing a fictitious surface charge density. (Note that in the local limit of the 
surface plasmon problem considered so far, this situation occurs with a real surface
charge.) GF work, however, with the Fuchs $p$-parameter for non-specular scattering
and use the condition
\begin{align}
\phi^{\rm M}(-z) = p \phi^{\rm M}(z)~.
\label{eq:pseudomodel_GMF_p_model_field_connection}
\end{align}
for the electric potential $\phi^{\rm M}$ in the metallic pseudo-medium. 
The non-definite left-right
symmetry of this potential (neither even nor odd)
requires additional fictitious stimuli that GF take
as a magnetic surface current and an electric surface dipole. For problems with
an externally applied field, also volume charges are allowed for that can be 
understood as generating the external field. For the vacuum pseudo-medium,
only a charge sheet is required.

The last condition fixes the surface current. Since GF try to avoid 
what they call an ``unphysical charge accumulation'', they set the surface current 
to zero \cite{GMF_cond_Surfaces_non_specular}
\begin{equation}
J_z(0^+) = 0.
\label{eq:pseudomodel_GMF_zero_surface_current}
\end{equation}
Only antisymmetric stimuli contribute to the surface current so that this
condition fixes a relation between the magnetic surface current and the 
electric surface dipole. The latter can be understood as a particular
choice of Zaremba's correction to the Fuchs boundary condition, 
that yielded the function $(1-p)A$ of 
Eqs.\,(\ref{eq:advanced_surface_scatt_boundary_cond}, \ref{eq:advanced_surface_scatt_boundary_cond_A}).

\subsection{Complex plasmon dispersion relation}
\label{subsec:disp_rel_HDM_GMF}

To calculate the surface plasmon dispersion, 
GF consider a collisionless hydrodynamic model
with longitudinal and transverse dielectric functions
[see Eq.\,(\ref{eq:HDM-dielectric-function})]
\begin{equation}
\epsilon_{\mu}(K, \omega) 
= 1 - \frac{\omega^2_{\rm p}}{\omega^2 - \beta^2_{\mu}K^2}
\quad \text{with} \quad
\mu = {\rm L,T}
\label{eq:pseudomodel_GMF_hydro_diel_fctn}
\end{equation}
Here, $\beta_{\rm L}$ and $\beta_{\rm T}$ are the corresponding sound velocities. 
Their values can be fixed by expanding Eq.\,(\ref{eq:pseudomodel_GMF_hydro_diel_fctn}) 
to second order in $K\beta_{\mu} / \omega$ and comparing to the same expansion 
of the Lindhard dielectric functions:
\begin{align}
\beta^2_{\rm L} = \frac{3}{5}v^2_{\rm F}
\qquad\text{and}\qquad
\beta^2_{\rm T} = \frac{1}{3}v^2_{\rm F}~. 
\end{align}
The fields generated by the fictitious stimuli in the two pseudo-media are
then constrained by the continuity relations of Maxwell's equations,
as well as Eqs.\,(\ref{eq:pseudomodel_GMF_p_model_field_connection}, \ref{eq:pseudomodel_GMF_zero_surface_current}). This determines all stimuli
and yields, after expansion for small $k \beta_\mu \ll \omega_{\rm s}$,
the dispersion relation
\begin{align}
\omega(k) = \omega_{\rm s} + \frac{k\beta_{\rm L}}{2} - \mathrm{i}\frac{1-p}{4}k\beta_{\rm T}
\label{eq:pseudomodel_GMF_longwave_disp_rel}
\end{align} 
The surface plasmon damping that appears here is attributed by GF to 
the mixing of surface and bulk modes\cite{GMF_intro_theory_solid_surfaces},
somewhat similar to the coupling between surface and bulk charges in
Sec.\,\ref{subsubsec:phenom_finite_surf_curr_HH_ABC}.
It is proportional to the diffusely scattering probability $1-p$. 
A damping arising from the bulk metal (for example due to Landau processes)
would require a modification of the dielectric 
functions~(\ref{eq:pseudomodel_GMF_hydro_diel_fctn}).

\begin{table*}
\caption{Comparison between surface plasmon dispersion relations for different models:
local dielectric (LD), specular reflection (SR), dielectric approximation (DA),
and the pseudo-model (GF). Approaches that may be used with any bulk dielectric
function are evaluated using the 
hydrodynamic response~(\ref{eq:pseudomodel_GMF_hydro_diel_fctn}).
L/T -- longitudinal/transverse fields with characteristic sound speeds
$\beta_{\rm L, T}$; 
$\omega_{\rm s}$ -- surface plasmon resonance in the long-wavelength limit;
$p \in [0,1]$ -- fraction of specularly scattered electrons (Fuchs parameter).}
\label{tab:embedding_GMF_HDM_SRM}\begin{ruledtabular}
\begin{tabular}{l|cccc}
	Model & LD & SR & DA & GF  
	\\ \toprule 
	Symmetry breaking? & no & yes & no & for $p\ne0$ \\
	L/T & L = T & L & L & L + T\\
	Surface scattering & not resolved & specular & diffuse & partially diffuse 
	\\
	Dispersion relation $\omega(k)$ & 
	$\omega_{\rm s}$ & 
	$\omega_{\rm s} + \frac{1}{2} k\beta_{\mathrm{L}}$ & 
	{$\omega_{\rm s} + \frac{1}{2} k 
		\left(\frac{1}{2} \beta_{\mathrm{L}} - i \beta_{\mathrm{L}}\right)$} & 
	{$\omega_{\rm s} + \frac{1}{2} k
		\left( \beta_{\mathrm{L}} - \frac{\mathrm{i}}{2}(1 - p)\beta_{\mathrm{T}}\right)$}
		\end{tabular}	
\end{ruledtabular}	
\end{table*}

A comparison to other results for the surface plasmon dispersion is shown
in Table\,\ref{tab:embedding_GMF_HDM_SRM}. 
If we specialize the case of pure specular scattering at the surface (specular
reflection model, $p = 1$), 
GF find that all fictitious stimuli except the charge sheet vanish (as Deng
also mentions in Appendix~B of Ref.\,\onlinecite{Deng_2019}). The extended-medium
construction of the fields in that case reduces to the derivation of
Ritchie and Marusak \cite{Ritchie_Marusak}. 
{It is worth recalling that the specular approximation can be made for any
choice of bulk dielectric function, in contrast to the construction of 
Sec.\,4.3 in Ref.\,\onlinecite{Deng_2019}.}

\subsection{The dielectric approximation} 

Heinrichs has introduced the so-called dielectric approximation (DA) \cite{Heinrichs_PhysRevB.7.3487} to compute the surface plasmon dispersion.
Using a sharp surface model, he introduces the constitutive relation for
the displacement field
\begin{equation}
z > 0: \qquad
\bm{D}(z) = \int_{z'>0}\!dz' \, \epsilon(z - z',k,\omega) \bm{E}(z')\,,
\label{eq:Heinrichs-constitutive-relation}
\end{equation}
where $\epsilon$ is the bulk dielectric function. The presence of the surface
is only taken into account by cutting the integral off at $z' = 0$. 
Heinrichs acknowledges the formal analogy of this constitutive relation 
to Reuter and Sondheimer's treatment of the anomalous skin effect 
\cite{Reuter_Sondheimer_anomalous_Skineffect}
for diffuse reflection ($p = 0$), in particular in view
of Eq.\,(\ref{eq:pseudomodel_GMF_p_model_field_connection}).
This is an identification that he objects to, however, putting forward
the different physical situations
in the surface plasmon problem and the anomalous skin effect.

The dielectric approximation does have a problem with charge conservation,
however, which has also been noted in Ref.\,\onlinecite{Mead_1977}, for example.
Indeed, if we compute the surface current from 
Eq.\,(\ref{eq:Heinrichs-constitutive-relation}), we get in general a nonzero
result (except for specific choices of the medium field $\bm{E}(z')$), so 
that one should deal with a surface charge 
[Eq.\,(\ref{eq:surface_charge_conservation_no_surface_divergence}), 
see also Eq.\,(40) of 
Ref.\,\onlinecite{Heinrichs_PhysRevB.7.3487}].
This is the reason why Heinrichs' long-wave dispersion relation
(see Table~\ref{tab:embedding_GMF_HDM_SRM})
shows a different slope,
if we compare to 
Eq.\,(\ref{eq:pseudomodel_GMF_longwave_disp_rel}) of the GF pseudomodel.
Heinrichs' result hence also differs from Ritchie and Marusak's specular
reflection model (which by symmetry prevents a surface charge,
see Eq.\,(\ref{eq:HDM-spdr-small-k})
and Eq.\,(26) of Ref.\,\onlinecite{Heinrichs_PhysRevB.7.3487}).

\section{\label{sec:conclusion}Conclusion}

We have tried in this paper to provide a transparent review of the historical
work on the plasmon dispersion relation at a (sharp) metallic surface, in order to
put the recent series of papers by Deng
\cite{Universal_Self_Amplification_Deng_et_al_2017,
Possible_Instability_Deng_2017,
Deng_2017b_metal_film, Deng_2019,
2017arXiv170603404D, 
2018arXiv180608308D,
Deng_2020a} into perspective. Deng advocates a different viewpoint
on the problem which is centered on the dynamics of charges and currents
rather than the electromagnetic field. The behavior of electrons at the surface
and in the sub-surface region plays a central role, both for the real part
(linear dispersion) and the imaginary part (damping vs.\ instability) of 
the surface plasmon frequency. We have emphasized how to implement the
conservation of charges in these processes, a basic task that does not seem
to be fully addressed in Deng's work. Already his starting point, the equation
of continuity, contains an unphysical loss term attributed to ``charge 
relaxation''. It turns out that a careful solution of the surface plasmon
problem does not show any instability. There is a small fraction of electrons
that ``flow uphill'' the electric potential at a diffusely scattering surface
and give energy to the electromagnetic field.
Their contribution is overwhelmed by other loss channels, however. 
Deng's alternative reasoning based on energy conservation is shown to be
technically flawed because the claimed contribution from the surface
current (the electric current density extrapolated to the inner metallic surface) 
actually does not contribute. Throughout the calculations, we have tried
to stay close to Deng's approach. Another technical issue that we found is
the nontrivial convergence of cosine-transformed fields at large momentum.
It cannot be excluded that despite claims to the contrary, Deng's results
actually depend on the cutoff momentum.

\acknowledgments

The work by CH was supported by the \emph{Deutsche Forschungsgemeinschaft}
within the \emph{DIP} program (grants no. Fo 703/2-1 and Schm 1049/7-1).

\appendix

\section{Checking the results of Deng}
\label{a:check-Deng-gsgb}

\subsection{Electric potential}
\label{as:cos-and-exp-integrands}

Deng \cite{Possible_Instability_Deng_2017,Deng_2019}
uses the cosine transform~(\ref{eq:definition_cosine_transform}) to
represent the charge density. 
The resulting equations for the electric field
are Eq.\,(9) in Ref.\,\onlinecite{Possible_Instability_Deng_2017}
and
Eqs.(17,18) in Ref.\,\onlinecite{Deng_2019} (adapted to a vacuum$|$metal interface):
\begin{align}
E_x(z) &= -{\rm i} \int_{0}^{\infty}\!{\rm d}q\,
\frac{ 4 k \rho_q }{ K^2 } \left( 2 \cos q z - {\rm e}^{ - k z } \right)
\label{eq:Deng-Ex-dq-integral}
\\
\text{and}~~~E_z(z) &= \int_{0}^{\infty}\!{\rm d}q\,
\frac{ 4 \rho_q }{ K^2 } \left( 2 q \sin q z - k \, {\rm e}^{ - k z } \right)
\label{eq:Deng-Ez-dq-integral}
\end{align}
with $K^2 = k^2 + q^2$. {Deng evaluates the two terms in the parentheses 
under the integrals separately when solving the
Boltzmann equation.} We want to illustrate here that they are
closely related. The key approximation is to assume that the charge
density is well localized on the scale $1/k$ on which the potential
varies (``charge sheet''). 
It is equivalent to replace $\rho_q \to \rho_{\rm s}$ and
to pull it out of the integrals. {As mentioned in the main text, this
approximation is also used by Deng, it is only the organization of the 
calculation that differs.}

The first term in Eq.\,(\ref{eq:Deng-Ex-dq-integral}) is
\begin{equation}
- {\rm i} \int_{0}^{\infty}\!{\rm d}q\,
\frac{ 4 k \rho_q }{ K^2 } 2 \cos q z
=
- {\rm i} \int_{-\infty}^{\infty}\!{\rm d}q\,
\frac{ 4 k \rho_q }{ q^2 + k^2 } \cos q z
\,,
\label{eq:evaluate-Ex}
\end{equation}
using an even extension of $\rho_q$ to $q < 0$
[see Ref.\,\onlinecite[after Eq.\,(31)]{Possible_Instability_Deng_2017}].
We apply the charge-sheet approximation and write the remaining integral as the real part of a
contour integral
\begin{align}
- 4 {\rm i} k \rho_{\rm s} \mathop{\rm Re}
\int_{-\infty}^{\infty}\!{\rm d}q\,
\frac{ {\rm e}^{ {\rm i} q z } }{ q^2 + k^2 } 
& =
- 4 {\rm i} k \rho_{\rm s} \mathop{\rm Re}
2\pi {\rm i} \frac{ {\rm e}^{- k z } }{ 2 {\rm i} k }
\nonumber\\
&=
- 4 \pi {\rm i} \rho_{\rm s}
{\rm e}^{ - k z }\,.
\label{eq:Ex-cos-term}
\end{align}
Assuming $z > 0$, the integration contour has been closed in the upper
half-plane, picking the residue at $q = {\rm i} k$. 

The other term in Eq.\,(\ref{eq:Deng-Ex-dq-integral}) becomes an elementary
integral for $\rho_q \to \rho_{\rm s}$, 
but could also be handled in the same way:
\begin{equation}
{\rm i} \rho_{\rm s} \int_{0}^{\infty}\!{\rm d}q\,
\frac{ 4 k }{ q^2 + k^2 } {\rm e}^{ - k z }
= 2 \pi {\rm i} \rho_{\rm s}
{\rm e}^{ - k z }
\,.
\label{eq:}
\end{equation}
This cancels half of the cosine term in Eq.\,(\ref{eq:Ex-cos-term}), giving
\begin{equation}
E_x(z) = - 2 \pi {\rm i} \rho_{\rm s}
{\rm e}^{ - k z }\,.
\label{eq:result-Ex}
\end{equation}
This agrees with Eq.\,(\ref{eq:electric_field_displaced_charge}) for $z > d \to 0$.

For the evaluation of $E_z$, one could re-use the previous result because
it gives, up to a factor $-ik$, the electric potential:
\begin{equation}
z > 0:\qquad
\phi(z) = \frac{ 2 \pi \rho_{\rm s} }{ k } {\rm e}^{ - k z }
\label{eq:result-potential}
\end{equation}
in agreement with~(\ref{eq:electric_potential_displaced_charge}).
Repeating the calculation with contour integrals is instructive,
however, because it illustrates how the exponential ${\rm e}^{ {\rm i} q z }$
regularizes the integral in the UV:
\begin{align}
4 \rho_{\rm s}
\int_{0}^{\infty}\!{\rm d}q\,
\frac{ 2 q \sin q z }{ q^2 + k^2 }
&= 
4 \rho_{\rm s}
\mathop{\rm Im}
\int_{-\infty}^{\infty}\!{\rm d}q\,
\frac{ q \, {\rm e}^{ {\rm i} q z } }{ q^2 + k^2 }\,.
\label{eq:}
\end{align}
Closing the contour for $z > 0$ in the upper half-plane and picking
the pole at $q = {\rm i} k$, one gets
\begin{equation}
4 \rho_{\rm s}
\int_{0}^{\infty}\!{\rm d}q\,
\frac{ 2 q \sin q z }{ q^2 + k^2 }
=
4 \pi \rho_{\rm s} \, {\rm e}^{ - k z }\,.
\label{eq:Ez-first-term}
\end{equation}
This result illustrates the fallacies of taking the limit $z \to 0$
too early (under the integral) because of the poor UV convergence
(see the discussion in Sec.\,\ref{sss:discussion-boundary-terms} for
examples from Deng's papers).
One half of the expression~(\ref{eq:Ez-first-term}) is subtracted 
by the second term in Eq.\,(\ref{eq:Deng-Ez-dq-integral}) so that we 
finally have
\begin{equation}
E_z(z) = 2 \pi \rho_{\rm s} \, {\rm e}^{ - k z }
\label{eq:result-Ex}
\end{equation}
in agreement with the potential~(\ref{eq:result-potential}).

\subsection{Distribution function}
\label{a:check-gb-gs}

We re-calculate here the distribution function of the semiclassical model,
using the formulas of Deng's papers. The idea is similar to the preceding
Appendix: by evaluating the $q$-integrations first, we get explicit
results that avoid convergence problems at large $q$. 
We use the notation $g_{b,s}$ of Deng 
[Eqs.(19--22) in Ref.\,\onlinecite{Possible_Instability_Deng_2017}
and Eqs.(32--35) in Ref.\,\onlinecite{Deng_2019}]
because we are going to see that
the splitting into ``bulk'' and ``surface'' is not unique.
\footnote{An obvious typo appears in Eq.\,(32) of Ref.\,\onlinecite{Deng_2019}
where the expression ``$2{\rm i}F - \sin(qz)$'' should be understood as
$2{\rm i}F_- \sin(qz)$.
In Eqs.(6, 7) and Eqs.(A6--A8) of
Ref.\,\onlinecite{Universal_Self_Amplification_Deng_et_al_2017}
and in Eq.\,(12) of
Ref.\,\onlinecite{Deng_2017b_metal_film},
a slightly different expression is given for the terms 
$- 2(1-p) F_+$ 
and
$- p F_0( \bm{k}, \bar{\omega}, \bm{v}_- )$ 
in Eq.\,(\ref{eq:Deng-g-surface}).
In Ref.\,\onlinecite{Deng_2017b_metal_film}, we have taken the thick-film 
limit $d \to \infty$. These may be typos.
}
\begin{align}
g_{b}( \bm{v}, z ) &= - e f'_0 \int_0^{\infty}\! dq
\frac{ 4 \rho_q }{ q^2 + k^2 } 
\left(
2 F_+ \cos(qz) 
\right.
\nonumber\\
& \qquad \left. {}
+ 2 {\rm i}F_- \sin(qz) - F_0 \, {\rm e}^{-k z}
\right)
\label{eq:Deng-g-bulk}
\\
g_{s}( \bm{v}, z ) &= - e f'_0 \Theta( v_z) 
\, {\rm e}^{ -\eta z } \kern-0.5ex \int_0^{\infty}\! dq
\frac{ 4 \rho_q }{ q^2 + k^2 } 
\left[
F_0( \bm{k}, \bar{\omega}, \bm{v} )
\right.
\nonumber\\
& \kern-2ex \left. {}
- p F_0( \bm{k}, \bar{\omega}, \bm{v}_- )
- 2(1-p) F_+( \bm{K}, \bar{\omega}, \bm{v} )
\right]\,.
\label{eq:Deng-g-surface}
\end{align}
Here, the abbreviations
\begin{align}
F_0 &= F_0( \bm{k}, \bar{\omega}, \bm{v} ) = 
\frac{ \bm{k} \cdot \bm{v} }{ \bar{\omega} - \bm{k} \cdot \bm{v} }
\,,\qquad
\bm{k} = ( k, 0, {\rm i}k )\,,
\label{eq:def-Deng-F0}
\\
F_\pm &= F_\pm( \bm{K}, \bar{\omega}, \bm{v} )
= \frac{1}{2} \left[
\frac{ \bm{K} \cdot \bm{v} }{ \bar{\omega} - \bm{K} \cdot \bm{v} }
\pm
\frac{ \bm{K} \cdot \bm{v}_- }{ \bar{\omega} - \bm{K} \cdot \bm{v}_- }
\right]\,,
\nonumber\\
&\qquad
\bm{K} = ( k, 0, q )
\label{eq:def-Deng-Fplus-minus}
\end{align}
are used. The function
$F_+$ ($F_-$) is even (odd) in the product variable $q v_z$, respectively. 
We evaluate
the $dq$-integral in the charge-sheet approximation $\rho_q \to \rho_{\rm s}$. 
The simplest case is
\begin{equation}
4 \rho_{\rm s} \int_{0}^{\infty}\!dq 
\frac{ (- F_0) \,{\rm e}^{ - k z } }{ q^2 + k^2 }
=
- 2\pi \rho_{\rm s} 
F_0 \frac{ {\rm e}^{ - k z } }{ k }\,.
\label{eq:Deng-gb-F0}
\end{equation}
The other integral in $g_{\rm b}$ is extended to the entire real
axis in the form
\begin{align}
& 4 \rho_{\rm s} \int_{-\infty}^{\infty}\!dq \frac{ 
F_+ {\rm e}^{ {\rm i} q z } +
F_- {\rm e}^{ {\rm i} q z } }{ q^2 + k^2 }
\nonumber\\
& \qquad
=
4 \rho_{\rm s} \int_{-\infty}^{\infty}\!dq \frac{ {\rm e}^{ {\rm i} q z} }{ q^2 + k^2 }
\frac{ \bm{K} \cdot \bm{v} }{ \bar{\omega} - \bm{K} \cdot \bm{v} }\,.
\label{eq:A16}
\end{align}
We look for poles in the upper half-plane and find $q = {\rm i} k$ and
$q v_z = \bar{\omega} - k v_x = {\rm i} v_z \eta$,
provided $v_z > 0$ (assuming $\mathop{\rm Im}\bar{\omega} > 0$).
Working out the residues, Eq.\,(\ref{eq:A16}) turns into
\begin{align}
& =
4 \pi \rho_{\rm s} \frac{ {\rm e}^{ -k z} }{ k }
F_0( \bm{k}, \bar{\omega}, \bm{v} )
-
4 \pi \rho_{\rm s} \Theta(v_z)  \frac{ 2 {\rm i} \bar{\omega} }{ v_z }
\frac{ {\rm e}^{ - \eta z} }{ k^2 - \eta^2 }
\label{eq:Deng-gb-Fplus-minus}
\end{align}
Added to Eq.\,(\ref{eq:Deng-gb-F0}), 
the first term changes the sign of the latter. 
Note the second term here
whose structure is interpreted by Deng as
describing electrons that move ballistically away from the surface ($v_z>0$). 
In the calculation presented in Sec.\,\ref{subsec:Boltzmann-approximate-solution},
where the $dq$-integral is performed first within the real-space representation
of the electric field
[see Eq.\,(\ref{eq:electric_field_displaced_charge})],
this term is missing from the bulk distribution function $f_{\rm b}$.
Using the identity~(\ref{eq:identity_distribution_functions}),
a similar term appears only in the surface distribution $f_{\rm s}$
[see Eqs.\,(\ref{eq:metal_volume_distribution_fctns})].

The first two terms of $g_{\rm s}$ give
\begin{align}
& \rho_{\rm s} \Theta( v_z) 
\, {\rm e}^{ -\eta z } \kern-0.5ex \int_0^{\infty}\! dq
\frac{ 4 }{ q^2 + k^2 } 
\left[
F_0( \bm{k}, \bar{\omega}, \bm{v} )
- p F_0( \bm{k}, \bar{\omega}, \bm{v}_- )
\right]
\nonumber\\
& =
2\pi \rho_{\rm s} \Theta( v_z) \frac{ {\rm e}^{ -\eta z } }{ k }
\left[
F_0( \bm{k}, \bar{\omega}, \bm{v} )
- p F_0( \bm{k}, \bar{\omega}, \bm{v}_- )
\right]
\label{eq:Deng-gs-F0}
\end{align}

Finally, we evaluate the integral
\begin{align}
& - 2(1-p) \rho_{\rm s} \Theta( v_z)
\, {\rm e}^{ -\eta z } \kern-0.5ex \int_0^{\infty}\! dq
\frac{ 4  }{ q^2 + k^2 } 
F_+( \bm{K}, \bar{\omega}, \bm{v} )
\nonumber\\
& =
- (1-p) \rho_{\rm s} \Theta( v_z)
\, {\rm e}^{ -\eta z } \kern-0.5ex \int_{-\infty}^{\infty}\! dq
\frac{ 4 }{ q^2 + k^2 } 
F_+( \bm{K}, \bar{\omega}, \bm{v} )
\label{eq:A19}
\end{align}
and make it convergent by inserting ${\rm e}^{ {\rm i} q d}$ 
with $d \downarrow 0$.
We close in the upper half-plane and find residues at $q = {\rm i} k$
and {$q = {\rm i} \eta$}
from the first summand in $F_+$ only (it would be
the second one if we had used
${\rm e}^{ -{\rm i} q d}$ and closed in the lower half-plane, giving
the same result). We get for Eq.\,(\ref{eq:A19})
\begin{align}
&= 2\pi (1 - p)\rho_{\rm s} \Theta( v_z) {\rm e}^{ -\eta z } 
\label{eq:Deng-gs-Fplus}
\\
& \hphantom{=
}{} \times 
\left[ 
- \frac{ F_0( \bm{k}, \bar{\omega}, \bm{v} ) + F_0( \bm{k}, \bar{\omega}, \bm{v}_- ) }{ k }
+ \frac{ 2 {\rm i} \bar{\omega} }{ v_z (k^2 - \eta^2) }
\right]
\nonumber
\end{align}
Simple algebra gives the identity
\begin{equation}
\frac{ 2 {\rm i} \bar{\omega} }{ v_z(k^2 - \eta^2) }
=
\frac{ 
F_0( \bm{k}, \bar{\omega}, \bm{v} )
- F_0( \bm{k}, \bar{\omega}, \bm{v}_- )
}{ k } 
\label{eq:identity_distribution_functions}
\end{equation}
This leaves only the term $F_0( \bm{k}, \bar{\omega}, \bm{v}_- )$ 
in Eq.\,(\ref{eq:Deng-gs-Fplus}).

Using the identity~(\ref{eq:identity_distribution_functions})
in Eq.\,(\ref{eq:Deng-gb-Fplus-minus}) and
adding the results~(\ref{eq:Deng-gb-F0}, \ref{eq:Deng-gs-F0},
\ref{eq:Deng-gs-Fplus}), 
we eventually find
\begin{align}
\frac{ g_b + g_s }{ 2\pi e \rho_{\rm s} f'_0  } 
&=
- \frac{ {\rm e}^{ -k z} }{ k }
F_0( \bm{k}, \bar{\omega}, \bm{v} )
\label{eq:Deng-gb-gs-total}\\
& \hphantom{=
} 
{} + \Theta( v_z) \frac{ {\rm e}^{ -\eta z } }{ k }
\left[
F_0( \bm{k}, \bar{\omega}, \bm{v} )
- p F_0( \bm{k}, \bar{\omega}, \bm{v}_- )
\right]
\nonumber
\end{align}
It is easy to check that the two lines in 
Eq.\,(\ref{eq:Deng-gb-gs-total}) are identical to the first two lines
in Eq.\,(\ref{eq:metal_volume_distribution_fctns}). 
The cancellations and simplifications in this calculation are truly
remarkable. To better visualize the interrelations among the terms,
we display them in Table~\ref{t:Deng-gb-gs-terms}.


%

\end{document}